\documentclass[%
 aip,
 amsmath,amssymb,
 reprint,%
]{revtex4-1}

\usepackage{graphicx}
\usepackage{dcolumn}
\usepackage{bm}

\usepackage[utf8]{inputenc}
\usepackage[T1]{fontenc}
\usepackage{mathptmx}
\usepackage{etoolbox}

\usepackage{xcolor}

\newcommand\abinitio{{\it ab-initio} }

\makeatletter
\def\@email#1#2{%
 \endgroup
 \patchcmd{\titleblock@produce}
  {\frontmatter@RRAPformat}
  {\frontmatter@RRAPformat{\produce@RRAP{*#1\href{mailto:#2}{#2}}}\frontmatter@RRAPformat}
  {}{}
}%
\makeatother
\begin{document}

\preprint{AIP/123-QED}

\title
{Photocurrents, inverse Faraday effect and photospin Hall effect in Mn$_2$Au
}

\newcommand{\pgi}{Peter Gr\"unberg Institut and Institute for Advanced Simulation,
Forschungszentrum J\"ulich and JARA, 52425 J\"ulich, Germany}

\newcommand{\aachen}{Department of Physics, RWTH Aachen University, 52056 Aachen, Germany}

\newcommand{\mainz}{Institute of Physics, Johannes Gutenberg University Mainz, 55099 Mainz, Germany}

\author{M. Merte}
 \email{maximilian.merte@rwth-aachen.de}
 \affiliation{\pgi}
 \affiliation{\aachen}
 \affiliation{\mainz}

\author{F. Freimuth}%
 \email{f.freimuth@fz-juelich.de}
 \affiliation{\mainz }
 \affiliation{\pgi }

\author{D. Go}
 \affiliation{\pgi}

\author{T. Adamantopoulos}
 \affiliation{\pgi}
 \affiliation{\aachen}

\author{F. Lux}
 \affiliation{\mainz}
 
 \author{L. Plucinski}
 \affiliation{\pgi}
 
\author{O. Gomonay}
 \affiliation{\mainz}

\author{S. Bl\"ugel}
 \affiliation{\pgi}
 
\author{Y. Mokrousov}
 \email{y.mokrousov@fz-juelich.de}
 \affiliation{\pgi}
 \affiliation{\mainz}
 \homepage{http://www.Second.institution.edu/~Charlie.Author.}

\date{\today}

\begin{abstract}
Among antiferromagnetic materials, Mn$_2$Au is one of the most intensively studied, and it serves as a very popular platform for testing various ideas related to antiferromagnetic magnetotransport  and dynamics. Since recently, this material has also attracted considerable interest in the context of optical properties and optically-driven antiferromagnetic switching. In this work, we use first principles methods to explore the physics of charge  photocurrents, spin photocurrents and inverse Faraday effect in antiferromagnetic Mn$_2$Au. We predict the symmetry and magnitude of these effects, and speculate that they can be used for tracking the dynamics of staggered moments during switching. Our calculations reveal the emergence of large photocurrents of spin in collinear Mn$_2$Au, whose properties can be understood as a result of a non-linear optical version of spin Hall effect $-$ which we refer to as the {\it photospin Hall effect} $-$ encoded into the relation between the driving charge and resulting spin photocurrents.  Moreover, we suggest that even a very small canting in  Mn$_2$Au can give rise to colossal  spin photocurrents which are {\it chiral} in flavor. 
We conclude that the combination of staggered magnetization with the structural and electronic properties of this material results in a unique blend of prominent photocurrents, which makes Mn$_2$Au a unique platform for advanced optospintronics applications.

\end{abstract}

\maketitle



%
%
%
%
%
%
\section{Introduction}
\newcommand\todo[1]{\color{red}{ToDo: #1}\color{black}}

In recent years antiferromagnets (AFMs) have attracted enormous interest in the context of spintronics applications due to a set of  unique properties, which make them often more favorable when compared to conventional ferromagnetic materials\cite{jungwirth2016antiferromagnetic,jungwirth2018multiple,borders2016analogue}.
For example, random access memory devices based on antiferromagnetic materials have been already realized \cite{kosub2017purely}, and THz switching of antiferromagnetic memory devices via current-induced spin-torques was demonstrated \cite{olejnik2018terahertz,kosub2017purely}.
However, in order to gain a reliable control over antiferromagnetic THz dynamics one has to move away from conventional electronics which is not sufficiently fast for the purpose. To overcome this speed limitation the concept of all optical switching was proposed\cite{vahaplar2009ultrafast,oppeneer2016ab,barbeau2022nonequilibrium,oppeneer2021ultrafast}, were switching and read-out are performed optically, often with THz laser sources. As a consequence, optical manipulation and in particular the read-out of the N\'eel vector during optical dynamics in antiferromagnets becomes a matter of grave importance\cite{nvemec2018antiferromagnetic,kampfrath2011coherent}. 
In this context, the properties of optically generated photocurrents come to occupy a special place\cite{wunderlich2022ultrashort,fedianin2022selection,Farkas2022optical,Qiong2022photocurrent,ciccarelli2023helicity}.

Some of major promises of antiferromagnetic spintronics hinge on our ability to get a control over the properties and timescales of ultrafast antiferromagnetic dynamics triggered by optical excitations. In the context of antiferromagnetic spintronics, Mn$_2$Au has emerged as a very successful representative, as it can be easily fabricated, it allows for Néel type of spin-orbit torques which are linear in the field, and which can be used to switch the staggered magnetization in this material,\cite{zelezny2014relativistic,bodnar2018writing,meinert2018electrical} and it is a good metal with very high N\'eel temperature, which is key for pronounced magnetotransport and magneto-optical response at room temperature. Owing to its relative structural and chemical simplicity, in the field of antiferromagetism Mn$_2$Au has long become a guinea pig for testing novel concepts relying on robust antiferromagnetic order and emergent functionalities. Optically-driven dynamical properties of Mn$_2$Au have become a target of recent attention as well. For example, recently, an optical manipulation of magnetic order in Mn$_2$Au has been achieved by a combination of laser pulses and strain\cite{demsar2021optical,grigorev2022optically}. However, in order to move forward in pursuing optical implementations based on Mn$_2$Au, it is important to understand in detail the optical response of this material, in particular the behavior of photocurrents and inverse Faraday effect, in relation to its electronic and magnetic structure.

In this work, we use the {\it ab-initio}-based Keldysh formalism, that we have applied in the past to various materials, to address the photo-response characteristics of bulk Mn$_2$Au. In particular, we study the microscopic origin, symmetry and magnitude of charge photocurrents in this material as a function of laser frequency, light polarization, band filling, degree of disorder and N\'eel vector direction. We identify qualitatively different types of photocurrents and demonstrate that by carefully measuring the evolution of photocurrents it is possible to track optical magnetization dynamics of Mn$_2$Au. We also predict the magnitude and properties of the uniform and staggered laser-induced spin polarization, i.e. inverse Faraday effect. In addition, by considering the laser-generated currents of spin, we introduce the notion of a {\it photospin Hall effect} and photospin Hall angle, which characterizes the ratio between the transverse photo-induced currents of charge and spin. Finally, we demonstrate that breaking the symmetry in the system even by small canting of staggered moments unleashes colossal {\it chiral spin photocurrents} whose sign can be controlled by the sense of canting. Overall, we provide an important qualitative and quantitative understanding of photo-induced phenomena in Mn$_2$Au, which may prove vital not only for further integration of this materials into the optical realm, but generally for engineering optical properties of wider classes of antiferromagnets.       

The manuscript is structured as follows. In Sec.~\ref{section2} we outline assumed electronic structure and  used computational methods. Sec.~\ref{section3a}  discusses the dependence of charge photocurrents on the N\'eel vector orientation, while in Sec.~\ref{section3b} properties of charge photocurrents are discussed for a fixed N\'eel vector orientation along the magnetic easy axis. In Sec.~\ref{section4} we analyze  inverse Faraday effect and report a large staggered response on the Mn sublattices.  Sec.~\ref{section5} discusses laser induced spin currents with a proposed photospin Hall effect in  Sec.~\ref{section5a}. In Sec.~\ref{section6} we analyze the effect of canting. Sec.~\ref{section6a} discusses the emergence of chiral photocurrents in the canted scenario. In close analogy we report on large chiral spin photocurrents driven by chiral inverse Faraday effect in  Sec.~\ref{section6b}. The manuscript ends with conclusions.
%
%
%

\begin{figure}
    \centering
    \includegraphics[width=1.0\columnwidth]{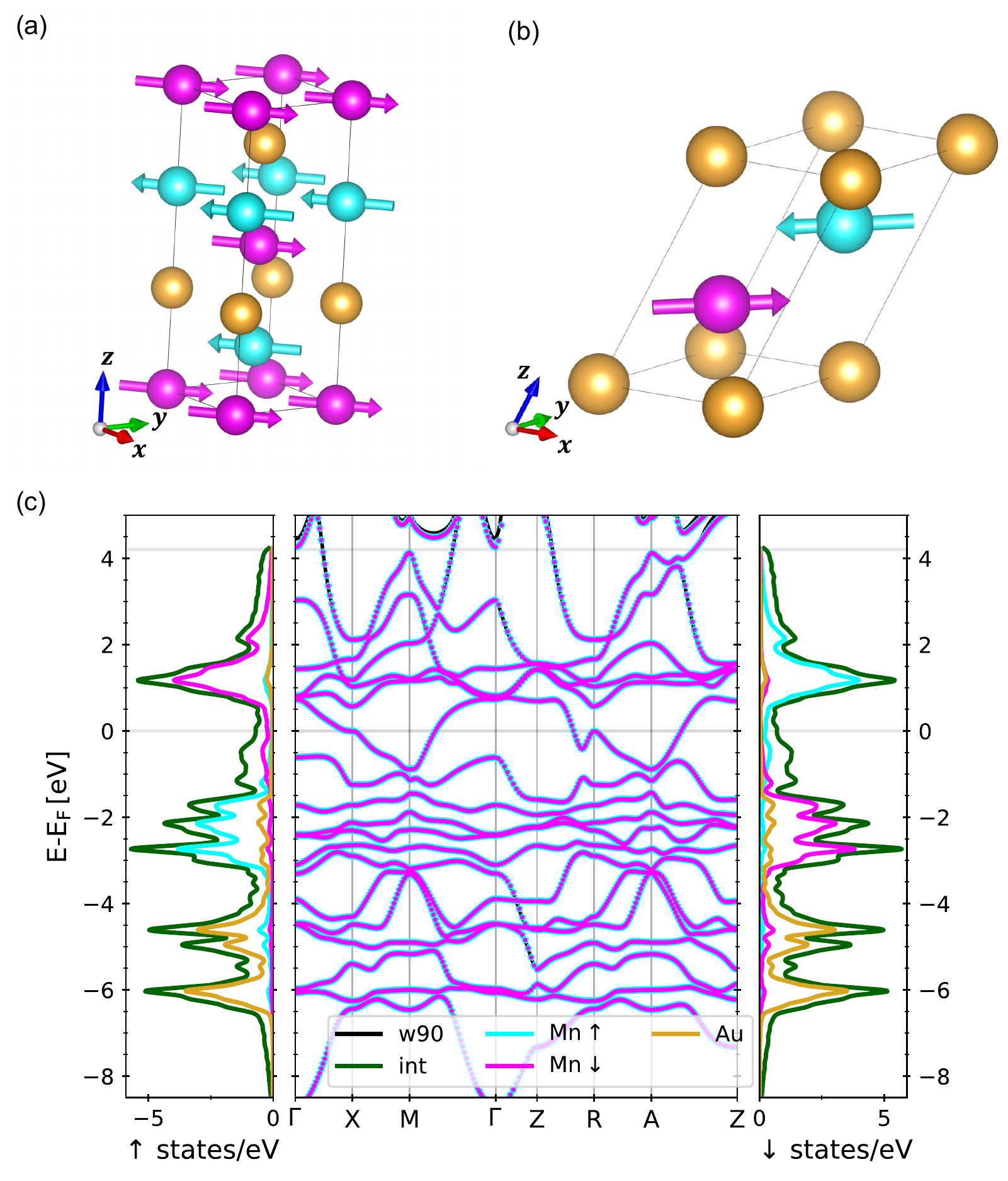}
    \caption[Mn2Au unit cell with staggered magnetization along the 110 direction (list  of figures descriptor).]{The unit cell of Mn$_2$Au with staggered magnetization along the [110]-direction. Cyan and magenta balls indicate Mn atoms with arrows depicting the magnetic moments. Golden balls represent  Au atoms. (a) Regular tetragonal unit cell hosting 6 atoms. (b) Reduced unit cell with  3 atoms. (c) Electronic properties of Mn$_2$Au with staggered magnetization along the [110]-direction. The left panel shows the ab-initio density of states (DOS) of spin down states decomposed into contributions from the interstitial region (green curve) and individual atomic sites Mn$\uparrow$, Mn$\downarrow$ and Au (cyan, magenta and golden curves). The right hand panel shows the DOS of the spin up states. In the center panel of (c) the ab-initio bandstructure (cyan and magenta dots for spin-up and spin-down states) is compared with the Wannier-interpolated bandstructure (black curve).  
    }
    \label{fig:mn2au_struct_comp}
\end{figure}

\section{Electronic structure and Computational Details}\label{section2}

The spatial inversion symmetry $\mathcal{P}$ is broken in Mn$_2$Au by the staggered magnetic moments on the Mn sublattices, see Fig.1(a,b), and thus only magnetic photocurrents are expected to survive. These are expected to strongly depend on the orientation and strength of the magnetic moments. On the other hand Mn$_2$Au possesses a combination of spatial inversion $\mathcal{P}$ with a time reversal operation $\mathcal{T}$, known as  $\mathcal{PT}$-symmetry. 
The characteristic $\mathcal{PT}$-symmetry of Mn$_2$Au is expected to suppress extrinsic contributions like the side-jump and skew-scattering contributions \cite{watanabe2020nonlinear} making Mn$_2$Au the ideal candidate to study the properties of pure magnetic photocurrents.

Mn$_2$Au has as tetragonal unit cell with space group \textit{I}4/\textit{mmm} \cite{wells1970structure}, see Fig.\ref{fig:mn2au_struct_comp}(a). The experimental lattice constant of $a=6.29 1$\,a.u.~and~$c=16.142$\,a.u.~were assumed for the calculations~\cite{wells1970structure}. The magnetization in the two Mn layers along the $z$-axis has collinear antiferromagnetic ordering \cite{khmelevskyi2008layered} and is assumed to be along the easy [110] direction, unless specified otherwise. In Fig.\ref{fig:mn2au_struct_comp}(a) we show the assumed structure. 
Electronic structure calculations were performed with the full-potential linearized augmented plane-wave code \texttt{FLEUR}\cite{fleur}. For self-consistent calculations we used a plane-wave cutoff of 
$4.0$\,a.u.$^{-1}$ and the total of 1728 $k$-points in the three-dimensional Brillouin zone. The plane wave cutoff for the potential (g$_{max}$) and exchange-correlation potential (g$_{max,xc}$) were set to 12.0 and 12.0\,a.u.$^{-1}$, respectively. The muffin-tin radii for Mn and Au atoms were set to 2.53\,a.u. and 2.60\,a.u., respectively. The nonrelativistic PBE~\cite{pbe} exchange-correlation functional was used. The spin-orbit coupling was included in second variation.


To reduce the numerical effort,  the size of the basis set was decreased by  half via using a reduced representation of the unit cell containing only 3 atoms, as shown in Fig.\ref{fig:mn2au_struct_comp}(b). For the initial projections of the maximally localized Wannier functions (MLWFs)\cite{wannier90}  $s$-,$p$-,$d$-orbitals were used for both Mn and Au atoms resulting in 54 Wannier functions for the case with 3 atoms in the unit cell.
In Fig.\ref{fig:mn2au_struct_comp}(c) we show the computed \abinitio bands in cyan and magenta for the spin up and down states respectively, while Wannier-interpolated bands are plotted in black. Within the frozen energy window, set to 13.6\,eV which is 4.2\,eV above the true Fermi level, the  interpolation shows a negligible deviation from the \abinitio band structure, which  allows to perform simulations of optical effects with frequencies up to $\hbar\omega=4.2$\,eV when evaluated at the true Fermi level. The electronic structure of Mn$_2$Au is manifestly metallic in nature. 
The density of states (DOS) for spin up (left) and spin down (right) states is shown in Fig.\ref{fig:mn2au_struct_comp}(c). 
In the relevant energy region for optical transitions of about 4\,eV around the Fermi level the largest contribution to the DOS comes from the interstial region and a similarly large contribution from the Mn atoms.  The states of Au  dominate the DOS only deeper into the valence bands (below $-$4\,eV). The DOS exhibits a large peak 
at $+$1.2\,eV above the Fermi level, and a group of similar peaks between $-$1.6\,eV and $-$3.2\,eV. 
The two Mn DOS (red and blue) undergo a visible spin flip when crossing from the group of peaks below the Fermi level to the region above it. The corresponding accumulation of states in these regions of energy may be associated with regions of pronounced response in some of the cases, as discussed further. 

In our work, we investigate second-order optical effects: charge and spin photocurrents, as well as non-equilibrium optically-induced spin density, known as the inverse Faraday effect\cite{ziel1965IFE,tatara2011inverse,oppeneer2016ab,freimuth2016torques}. Namely, we calculate the spin accumulation, charge and spin photocurrents which arise at second order in the perturbing electric field of a continuous laser pulse of frequency $\omega$ by using the expressions which were previously derived by us in~Ref.~\onlinecite{freimuth2016torques} within the framework of the Keldysh formalism, and applied to various systems\cite{freimuth_2021,merte2021photocurrents,adamantopoulos2022laser}. Throughout this work the assumed intensity of the pulse was set to 10\,GW/cm$^2$, which corresponds to typical values of the fluence of the order of 0.5\,mJ/cm$^2$ for a 50\,fs laser pulse~\cite{Huisman_2016}. Since the laser intensity enters the formalism only as a prefactor, see expressions in Refs.~\onlinecite{freimuth2016torques,freimuth2017laserinducedarxiv}, the results can be converted to other intensity values easily. A 256$\times$256$\times$256 $k$-mesh has proved to be sufficient to obtain converged results throughout the paper. In order to describe the effect of disorder of the electronic states on the computed quantities, the method of constant lifetime broadening $\Gamma$, naturally included into to Keldysh formalism, was used\cite{kadanoff1962quantum}.

As discussed e.g. by Azpiroz {\it et al.}~\cite{azpiroz2018abinitio}, we introduce the photoconductivity tensor $\sigma_{ijk}$ such that the photocurrent $ J_{i}=\tfrac{a_{0}^{2} e I}{\hbar c}\left(\tfrac{\mathcal{E}_{\mathrm{H}}}{ \hbar \omega}\right)^{2} \operatorname{Im} \sum_{j k} \epsilon_{j} \epsilon_{k}^{*} \varphi_{i j k} $ in Eq.1 of Ref.~\onlinecite{merte2021photocurrents} $-$ arising in response to the field $\mathbf{E}(t)=\mathrm{Re}[E_0\boldsymbol{\varepsilon}e^{-i\omega t }]$ where $\boldsymbol{\varepsilon}$ is the light-polarization vector and $E_0$ is the amplitude
of the electric field $-$ is given by $J_i= \mathrm{Re} \sum_{jk} 2 \sigma_{ijk} E_j E^*_k$, similarly to the definition of Eq.7 in Ref.~\onlinecite{azpiroz2018abinitio}. By inserting\cite{freimuth2016torques} the laser intensity $I=\epsilon_0 c E^2_0/2$  into Eq.1 of Ref.~\onlinecite{merte2021photocurrents} it follows for the second order photoconductivity $\varphi_{ijk}$ that $ \sigma_{ijk} = i \frac{a^2_0 e \epsilon_0}{2 \hbar}  \frac{\varphi_{ijk}}{2}$. In the expressions above, $a_0$ is the Bohr's radius, $e$ is the elementary charge, $I$ is the intensity of the pulse, $\hbar$ is the reduced Planck constant, $c$ is the light velocity, $\mathcal{E}_H=e^2/(4\pi \epsilon_0 a_0)$ is the Hartree energy, and $\epsilon_j$ is the $j$'th component of the polarization vector of the pulse. The expression for the spin photocurrent $Q^{s}_{i}$ flowing in direction $i$ with spin polarization along axis $s$, is obtained by replacing one of the velocity operators $v_i$ in  an expression for $J_i$, with the operator of the spin velocity $\{v_i,\tau_s\}$, where $\tau_s$ is one of the Pauli matrices, and the prefactor $a_{0}^{2} e I / \hbar c$  with the prefactor $-a_{0}^{2} I / 4 c$. 
For quantifying the inverse Faraday effect, the expression for the photo-induced spin density $\delta S_{i}$ is obtained by replacing one of the velocity operators $v_i$ in an expression for $J_i$, with the operator of (local) spin, and the prefactor $a_{0}^{2} e I / \hbar c$  with the prefactor $-\hbar\, a^3_0  I/2c$.
In our work, a circularly polarised pulse propagating for example in the $z$ direction is described by $\boldsymbol{\epsilon}=(1,\lambda i,0)/\sqrt{2}$, where $\lambda=\pm 1$ is the helicity. Linearly polarised light along for example $x$ or $y$ axis is described by $\boldsymbol{\epsilon}=(1,0,0)$ and $\boldsymbol{\epsilon}=(0,1,0)$, respectively, while generalization to arbitrary axes is obvious. 

\section{Photocurrents of charge}\label{section3}

First, we focus on the properties of the photocurrents of charge in Mn$_2$Au. We start by discussing the dependence of the photocurrent on the orientation of the N\'eel vector within the $xy$-plane, then fix the N\'eel vector to the [110]-direction and, after analyzing symmetry properties of the $k$-resolved photocurrents, proceed with the discussion of the spectral properties of the photocurrents. We conclude our discussion by analyzing the influence of lifetime broadening and identify magnetic photocurrents by considering broadening values of positive and negative sign. For the sake of simplicity, below we will refer to currents arising in response to linearly and circularly polarized light as {\it linear} and {\it circular}, respectively. This should not be confusing keeping in mind that  both types of responses we compute are manifestly second order in the applied field. 

\subsection{N\'eel vector dependence of the charge photocurrents}\label{section3a}
\label{sec:NeelDep}

Since the inversion symmetry $\mathcal{P}$ in Mn$_2$Au is broken only by the magnetization on the Mn sublattices and not the crystal structure itself, the resulting photocurrents are expected to be of purely magnetic origin. The orientation of the magnetic photocurrent is thus expected to depend directly on the orientation of the N\'eel vector. Indeed, in a recent experiment the magnetic linear dichroism was found to have a distinct dependence on the relative orientation of the linearly polarized light and the N\'eel vector\cite{grigorev2021optical}. Further the nonlinear anomalous Hall effect was recently proposed to measure the N\'eel vector orientation \cite{shao2020nonlinear}. We first compute the photocurrents  $J_x$ and $J_y$ for the N\'eel vector lying in the $(001)$-plane along the $x$- and $y$-directions for the frequency of linearly polarized light of 1.55\,eV and a lifetime of 100\,meV.
%
We find that in case of the N\'eel vector aligned along $x$
the $J_y$ component shows the largest response of almost 90$\times 10^{10}$\,A/m$^2$ when the  polarization is along $y$, while for the light polarized along $x$ the current response switches sign and is reduced in amplitude to $\sim$70$\times 10^{10}$\,A/m$^2$. When the N\'eel vector is aligned along $y$ the situation is reversed:
as a result of the symmetry properties of the system the overall angular dependence with respect to the plane of light polarization is shifted by 90$^\circ$ between the two N\'eel vector orientations  with an additional sign reversal of the resulting current, yielding:  
\begin{equation}
\begin{aligned}
    J_y (\mathbf{N}\parallel x;\mathbf{\varepsilon}\parallel y) &= -  J_x (\mathbf{N}\parallel y;\mathbf{\varepsilon}\parallel x)  \\
    J_y (\mathbf{N}\parallel x;\mathbf{\varepsilon}\parallel x) &= -  J_x (\mathbf{N}\parallel y;\mathbf{\varepsilon}\parallel y). 
\end{aligned}
\end{equation}

\begin{figure}
    \centering
    \includegraphics[width=1.0\columnwidth]{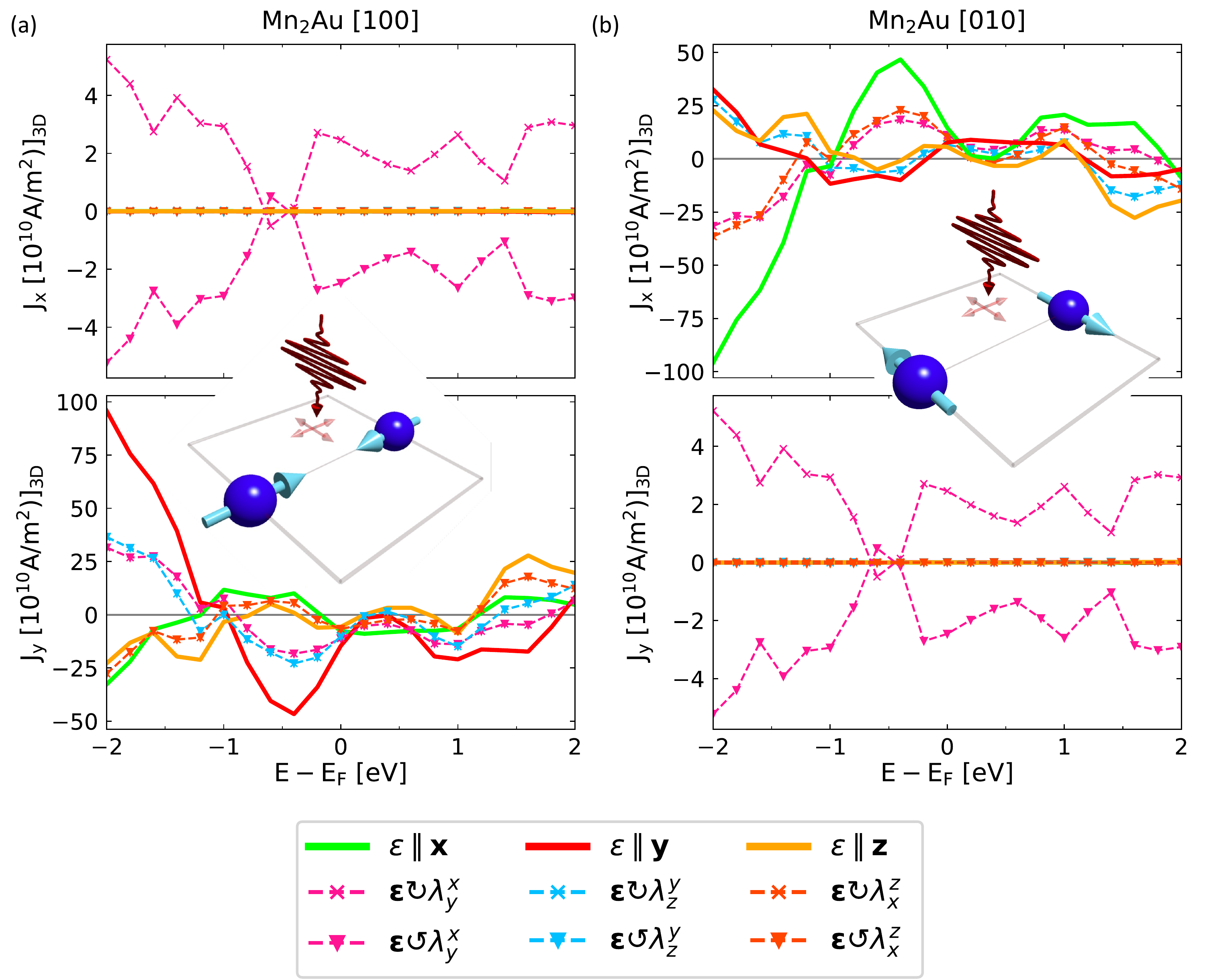}
    \caption[]{$J_x$ and $J_y$ components of the charge photocurrent in Mn$_2$Au for the direction of the N\'eel vector along $x$ (a) and $y$ (b), in response to red laser light with $\hbar \omega=1.55$\,eV as a function of band filling. The broadening was set to $\Gamma=100$\,meV.  }  
    \label{fig:mn2au_3at_100vs010}
\end{figure}

The fact that this symmetry property is generic is confirmed by calculations of the band filling dependence of the photocurrents, shown in Fig.\ref{fig:mn2au_3at_100vs010} for linearly and circularly polarized light of $\hbar\omega=1.55$\,eV and $\Gamma=100$\,meV for $\mathbf{N}$ along $x$ [left column, (a)] and $\mathbf{N}$ along $y$ [right column, (b)]. 
 In both scenarios the largest photocurrent is flowing perpendicular to the staggered magnetization, while the photocurrent parallel to $\mathbf{N}$ is an order of magnitude smaller. The perpendicular current is maximized for linearly polarized light, while parallel response is only driven by circularly polarized light. For the current perpendicular to the N\'eel vector finite response to circularly polarized light is sizeable, however it remains helicity-independent, and is shaped by responses to linearly polarized light. This indicates that the circular responses are just the averages of the  linear responses spanning the plane normal to the direction of incidence of circularly-polarized light (e.g.~the light blue curve in   Fig.\ref{fig:mn2au_3at_100vs010} is an average between the red and the yellow  curves). This behavior is characteristic of the {\it magnetic linear photogalvanic effect} (magnetic LPGE), while the circular responses can be interpreted as {\it magnetic circular photogalvanic effect} (magnetic CPGE). The data presented in Fig.\ref{fig:mn2au_3at_100vs010} shows that the correspondence between the two scenarios is universal over the entire energy range. 
\subsection{Photocurrents for N\'eel vector along the easy axis}\label{section3b}
Here, we study the photocurrents for the case of $\mathbf{N}$ along the easy [110] direction\cite{barthem2016easy}. Correspondingly, from now on till the end of the chapter, for brevity,  the responses will be discussed in terms of their projection onto $\mathbf{N}$  as $\parallel \mathbf{N}$, and as $\perp \mathbf{N}$ marking the projection onto the $[1\overline{1}0]$ 
direction. Since the [110] and [100] direction are the easy axes, and the $z$-direction is the hard axis\cite{barthem2016easy}, we will refer to components within the $xy$-plane as ``in-plane'' and $z$-components as ``out of plane'' responses. 
 The laser field polarization $\boldsymbol{\varepsilon}$ will be projected onto these directions as well, with linear polarized fields denoted by $\boldsymbol{\varepsilon} \parallel \mathbf{N}$, $\boldsymbol{\varepsilon} \perp \mathbf{N}$ and  $\boldsymbol{\varepsilon} \parallel z$.
 For circular polarized light rotating in the $xy$-plane we use $\boldsymbol{\varepsilon}\circlearrowleft\lambda^{y}_{x}$ for counter- and $\boldsymbol{\varepsilon}\circlearrowright\lambda^{y}_{x}$ for clockwise rotation. Besides light rotaing within the magnetic easy plane, we also consider circular polarized fields rotating in the plane spanned by $z$ with the two in-plane directions $\parallel\mathbf{N}$ and $\perp\mathbf{N}$, which will be denoted by $\boldsymbol{\varepsilon}\circlearrowright/\circlearrowleft\lambda^{z}_{\parallel\mathbf{N}}$ and $\boldsymbol{\varepsilon}\circlearrowright/\circlearrowleft\lambda^{z}_{\perp\mathbf{N}}$. 
 

\subsubsection{Band-resolved contributions.}
Before proceeding with a detailed analysis of the behavior of the integrated photocurrents, we would like to get an impression about the ``internal'' band-resolved and $k$-resolved constituents of these integrated quantities.  We start by scrutinizing the role of so-called {\it resonant transitions},\cite{adamantopoulos2022laser} i.e. contributions to the photocurrents which involve only two distinct band transitions, as opposed to a more general case where three distinct band transitions are involved (i.e.~non-resonant transitions). The resonant transitions can be seen as transitions where the initial and final state are the same. 
It is known that the LPGE is given by such resonant transitions, while the CPGE effects can only be described by non-resonant transitions.\cite{nagaosa2018photogalvanic,felser2019switchable}.
Indeed, as we have checked by explicit calculations, the magnetic LPGE currents $J_{\perp\mathbf{N}}$ discussed below 
are solely given by resonant transitions, while the magnetic CPGE currents $J_{\parallel\mathbf{N}}$ and  $J_{z}$ are originating solely in  non-resonant transitions. 

\begin{figure} 
    \centering
    \includegraphics[width=0.85\columnwidth]{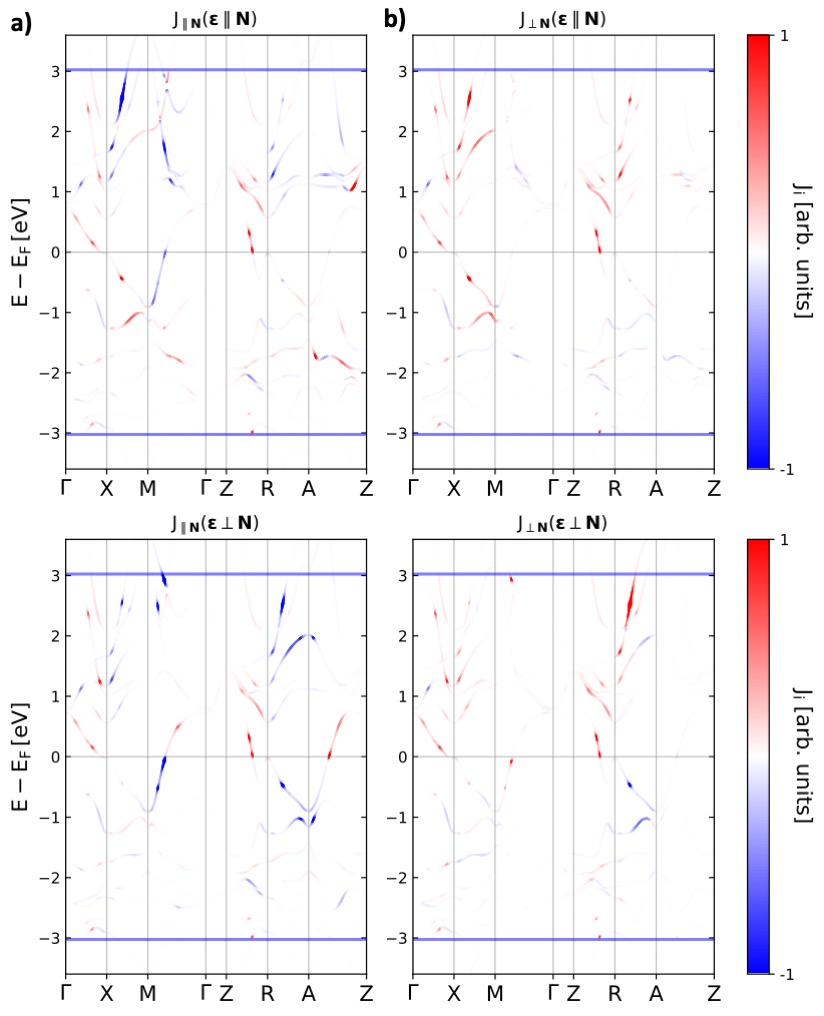}
    \caption[]{Resonant band contributions to the photocurrents plotted  along high symmetry lines in the Brillouin zone for (a) parallel, and (b) perpendicular photocurrent for linearly polarized light which is parallel (upper panel) or perpendicular (lower panel) to the N\'eel vector along [110].  Here, the laser frequency is $\hbar\omega=3.0$\,eV, $\Gamma=25$\,meV, the true Fermi level is indicated by the horizontal grey line, and the edges of the selected energy interval $[\mathrm{E_F}-\hbar\omega,\mathrm{E_F}+\hbar\omega]$ are indicated by blue lines. The strength of the response at a given $k$-point is indicated with band thickness and color.}
    \label{fig:mn2au_110_bandTrans_default}
\end{figure}

The resonant transitions can be naturally presented and discussed in a band-resolved manner in the reciprocal space~\cite{adamantopoulos2022laser}. In Fig.\ref{fig:mn2au_110_bandTrans_default} we show the  parallel and perpendicular photocurrents, given by resonant transitions, arising in response to blue $\hbar\omega=3.0$\,eV light for linear polarization parallel and perpendicular to the N\'eel vector. 
 As discussed above, the parallel component of the photocurrent integrates to zero over the Brillouin zone (BZ) with the perpendicular component being dominant. However, large resonant contributions are visible not only in the perpendicular scenario, shown in Fig.\ref{fig:mn2au_110_bandTrans_default}(b), but also in case of the parallel component, shown in Fig.\ref{fig:mn2au_110_bandTrans_default}(a). 
 
 Let us first compare the two scenarios for light polarized along the N\'eel vector, $\boldsymbol{\varepsilon}\parallel\mathbf{N}$ (upper panel of Fig.\ref{fig:mn2au_110_bandTrans_default}). Clearly, at a given $k$-point electronic transitions occur between bands which are separated by the frequency of 3.0\,eV. For example, a prominent transition takes place  between X and M points from $-$0.3\,eV to $+$2.7\,eV,  visible both in (a) and (b). In case of the nonvanishing perpendicular current, both energy regions contribute positive values, while in the case of parallel current the conduction and valence bands provide response which is opposite in sign. In other regions of the BZ both currents are very similar in strength and sign (see for example transition along $\Gamma$X or ZR).
 Along RA the transitions are of comparable strength but of the opposite sign between two cases. The regions from M to $\Gamma$ and from A to Z host much more pronounced contributions to  $J_{\parallel\mathbf{N}}$. 
 
 In case of linearly polarized light with polarization perpendicular to the N\'eel vector (lower pahel of Fig.\ref{fig:mn2au_110_bandTrans_default}) the distribution of significantly contributing transitions is notably different. Especially the transitions taking place along ZRA are dominant. Another prominent contribution is visible along M$\Gamma$ for both components of the current. 
 But overall, the transitions ignited by  linearly polarized light are visible along the high symmetry lines for both cases of $J_{\parallel\mathbf{N}}$ and $J_{\perp\mathbf{N}}$ often at the same $k$-points with only difference in the sign and strength of the transition. Transitions do take place also in case of  the symmetry-suppressed response $J_{\parallel\mathbf{N}}$, which suggests that they are cancelled by transitions of opposite sign in other regions of the BZ. 
 
\begin{figure}
    \centering
    \includegraphics[width=1.0\columnwidth]{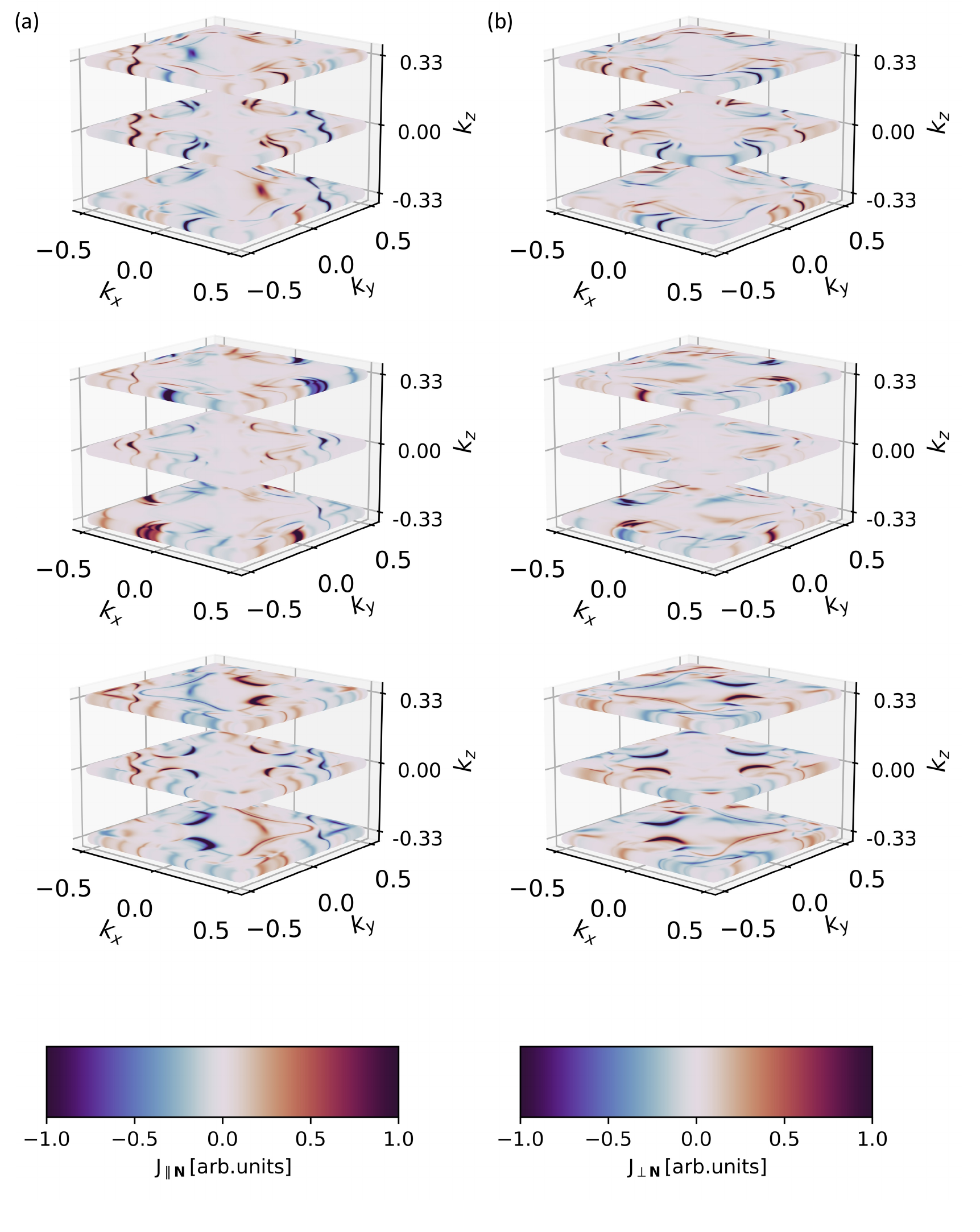}
    \caption[]{Photocurrents resolved in the Brillouin zone for blue laser light of $\hbar\omega=\,$3.0 eV at $\Gamma=\,$25 meV, in different $k_z$-layers. The left column shows the parallel photocurrent $J_{\parallel \mathbf{N}}$, the right column the perpendicular photocurrent $J_{\perp \mathbf{N}}$. The columns show different orientations of the linearly polarized light, in the first column light is polarized along the N\'eel vector $\boldsymbol{\varepsilon} \parallel \mathbf{N}$, the second shows the light polarization in the (001) plane but perpendicular to the Neel vector $\boldsymbol{\varepsilon} \perp \mathbf{N}$ and the third column the response for light polarized linearly along $\boldsymbol{\varepsilon} \parallel z$. }
    \label{fig:mn2au_110_bzRES_Jpara_vs_Jperp}
\end{figure}

 To get a better impression over the distribution of resonant transitions in the BZ, we analyze the distribution of the photocurrents in three slices of the BZ, which are parallel to  $(001)$-plane: at $k_z=0$ and $k^0_z=\pm 0.25$ in relative units, see Fig.\ref{fig:mn2au_110_bzRES_Jpara_vs_Jperp},
 which shows the parallel (left column) and perpendicular (right column) photocurrent as contour plots over these slices. 
In case of the overall nonvanishing $J_{\perp\mathbf{N}}$ the distribution of transitions has a perfect mirror symmetry between the $+k_z^0$ and $-k_z^0$ slices, see right part of Fig.\ref{fig:mn2au_110_bzRES_Jpara_vs_Jperp}.  For $J_{\parallel\mathbf{N}}$, on the other hand, the transitions are not only mirrored but they also change sign as we change the slice from that at $+k_z^0$ to that at $-k_z^0$, thus integrating to zero over the entire BZ. The observation that while the integrated response is vanishing but the local contributions are pronounced suggests a possibility that the $k$-contributions can  promote an integral response to linearly polarized light in $J_{\parallel\mathbf{N}}$ upon strain, canting (discussed further), or by going to third-order in the field responses. 
\subsubsection{Frequency and band filling dependence.}

We next study the frequency dependence of $J_{\perp\mathbf{N}}$ currents. In order to compare the strength of the photocurrents to the values available in the literature, 
it is convenient to refer to the conductivity tensor as  defined in Sec.~\ref{section2}.   
We further project the conductivity tensor  onto the $[1\overline{1}0]$-direction via $\sigma_{\perp\mathbf{N}jk}=(\sigma_{xjk}-\sigma_{yjk})/\sqrt{2}$ and use the following rules: 
\begin{equation}
    \begin{aligned}
    \sigma_{\perp\mathbf{N}}(\varepsilon\parallel\mathbf{N})  &:=  \tfrac{1}{2} (\sigma_{\perp\mathbf{N}xx}+\sigma_{\perp\mathbf{N}yy})+\tfrac{1}{2}(\sigma_{\perp\mathbf{N}xy}+\sigma_{\perp\mathbf{N}yx}),    \\
    \sigma_{\perp\mathbf{N}}(\varepsilon\perp\mathbf{N})      &:=  \tfrac{1}{2} (\sigma_{\perp\mathbf{N}xx}+\sigma_{\perp\mathbf{N}yy})-\tfrac{1}{2}(\sigma_{\perp\mathbf{N}xy}+\sigma_{\perp\mathbf{N}yx}), \\
     \sigma_{\perp\mathbf{N}}(\varepsilon\parallel\mathbf{z}) &:=	\sigma_{\perp\mathbf{N}zz},
    \end{aligned}
    \label{eq:this_tensor}
\end{equation}
 so as to bring the conductivity into the shape compatible with the sense of linear polarization of the light.

\begin{figure} 
    \centering
    \includegraphics[width=0.8\columnwidth]{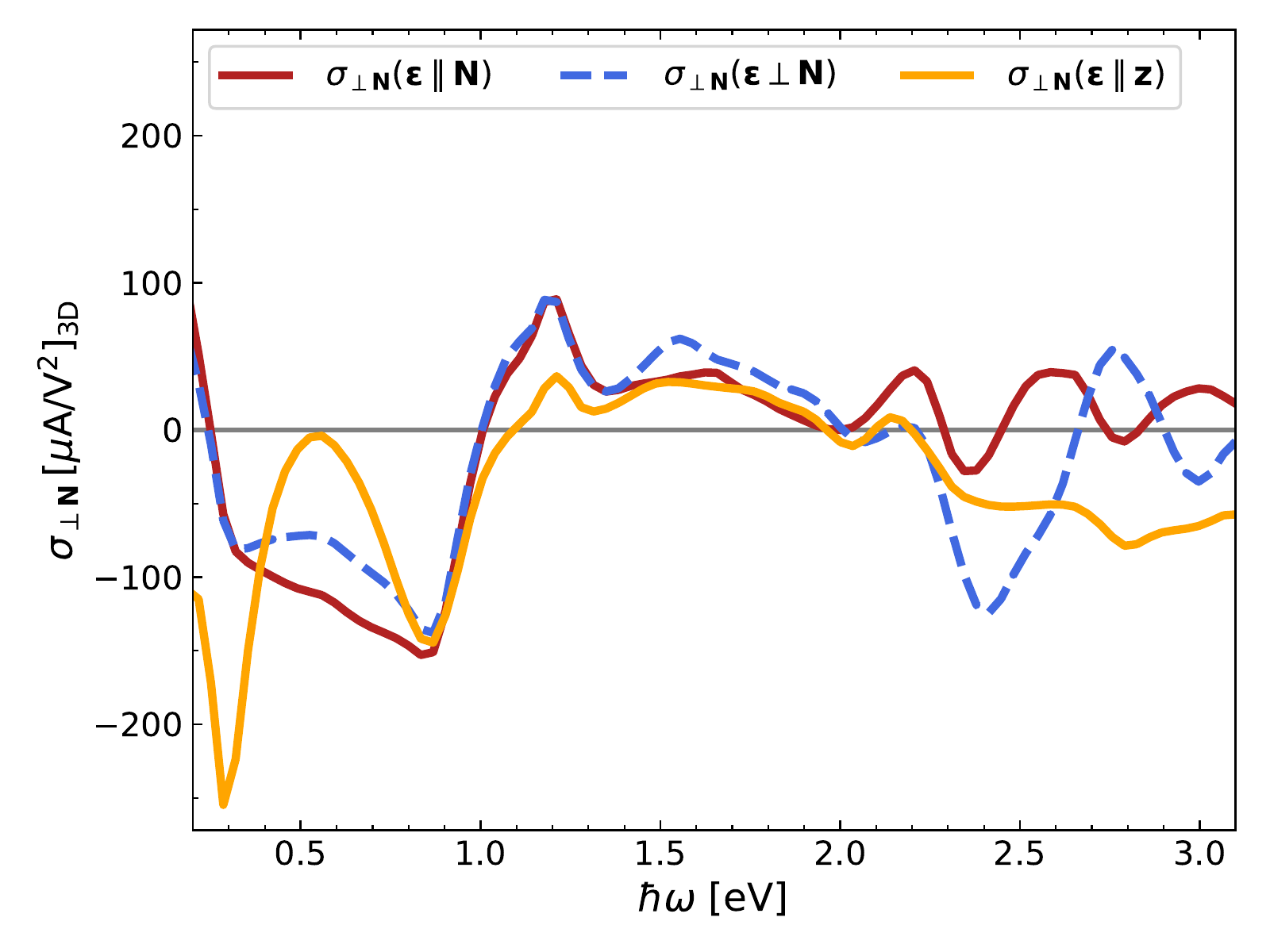}
    \includegraphics[width=0.8\columnwidth]{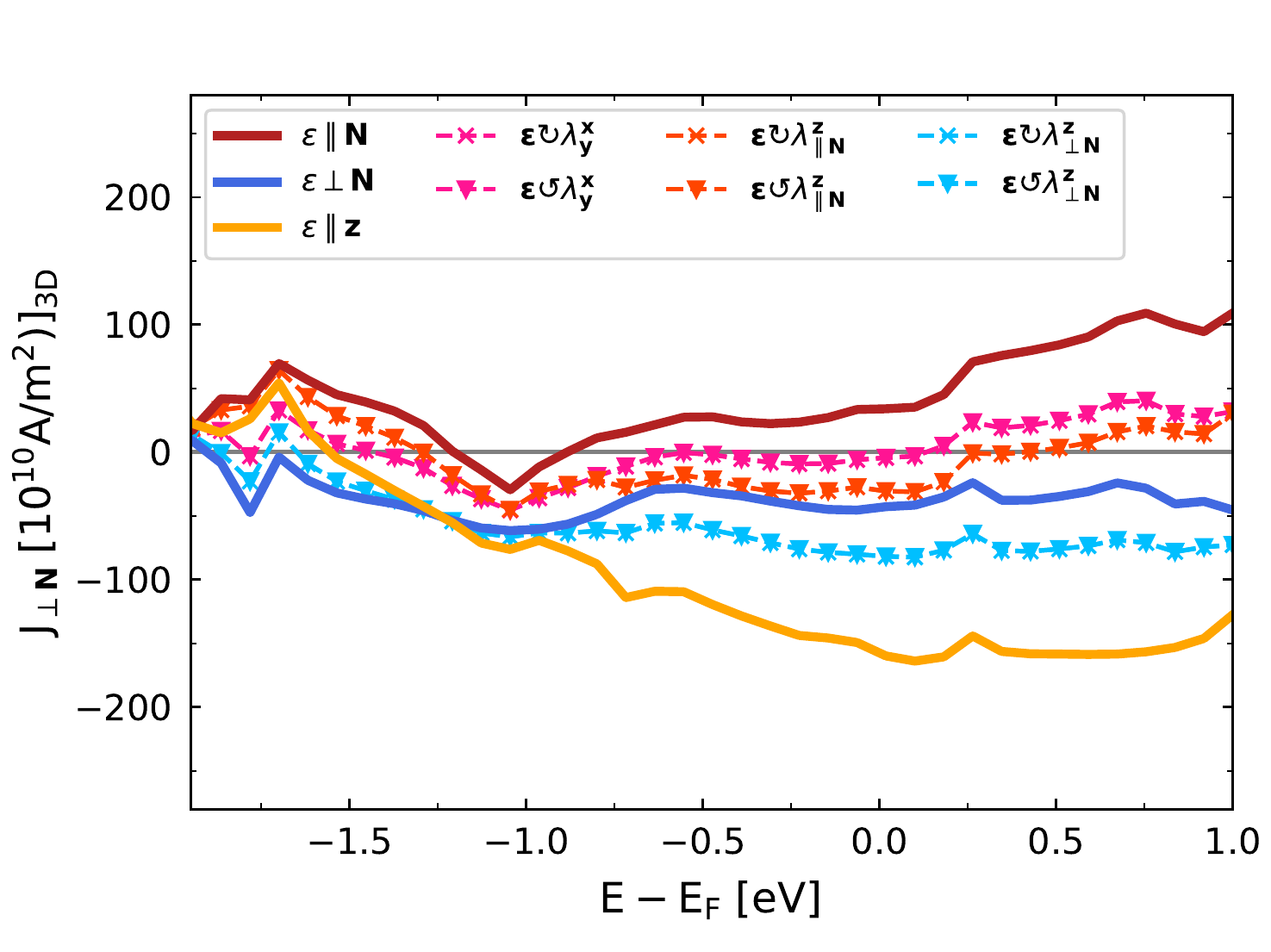}
    \caption[]{
    Properties of in-plane photocurrents perpendicular to the N\'eel vector, for $\hbar\omega=$\,3.00 eV and a broadening of $\Gamma=25$\,meV. (a) Frequency dependence of the conductivity tensor projected onto the $[1\overline{1}0]$ direction for different cases of linearly polarized light. The Fermi level was set to the true Fermi level of Mn$_2$Au.
    (b) Photocurrent $J_{\perp\mathbf{N}}$  as a function of band filling in Mn$_2$Au. 
    At the true Fermi level the largest photocurrent arises in response to a laser light linearly polarized along $z$. Note the difference in the band filling behavior to the case with another frequency and lifetime shown in Fig.\ref{fig:mn2au_3at_100vs010}. }
    \label{fig:mn2au_110_condTens_vs_hw}
\end{figure}

In Fig.\ref{fig:mn2au_110_condTens_vs_hw}(a) we present the frequency dependence of the conductivity tensor as defined in Eq.\eqref{eq:this_tensor} for different linear polarizations. 
We observe sizeable values of photoconductivity for all polarization flavors across the complete presented frequency range of $\hbar\omega \in [0.20,3.10]$\,eV. 
For laser frequencies below  $\hbar\omega=1\,$eV the conductivity is negative for all cases. At  $\hbar\omega=0.8\,$eV the conductivity exhibits a peak of about $- 150\,\mu$A/V$^2$ similar in magnitude for all polarizations. Another peak of about $+100\,\mu$A/V$^2$ is present at $\hbar\omega=1.20\,$eV if the laser light is linearly polarized within the $xy$-plane, while at the same time the $z$-linear  conductivity (orange line)  shows a much smaller peak at this frequency. These conductivity values are comparable to those reported in literature:~e.g.~Azpiroz {\it et al.}~\cite{azpiroz2018abinitio} reported a maximum conductivity of $\sim40\,\mu$A/V$^2$ for the bulk semiconductor GaAs, and more recently Zhang {\it et al.}~\cite{felser2019switchable} predicted a photoconductivity of almost $200\,\mu$A/V$^2$ in two-dimensional van-der-Waals antiferromagnet CrI$_3$. 

In the range of $\hbar\omega=1.20\,$eV up to about $\hbar\omega=2.00\,$eV the conductivity is positive for all polarizations. At $\hbar\omega=2.00\,$eV the two in-plane-polarized conductivities (blue and red lines) are  suppressed and start to oscillate with with frequency up to 3.1\,eV, remaining predominantly opposite in sign. This may prove useful: indeed, for example, the conductivity $\sigma_{\perp\mathbf{N}}$ can be switched in sign between identical in magnitude values of $\pm$25\,$\mu$A/V$^2$ at a laser frequency of $\hbar\omega=3.00\,$eV by simply rotating the linear polarization vector by 90$^\circ$ in the $xy$-plane. Owing to conveniently opposite sign of the two in-plane polarized responses at $\hbar\omega=3.00\,$, this frequency will be used as default for the remaining discussion.

By staying at $\hbar\omega=3.00$\,eV in Fig.\ref{fig:mn2au_110_condTens_vs_hw}(b) we present the dependence of $J_{\perp\mathbf{N}}$ on band filling. As discussed above, 
the circular responses (symbols) are helicity-independent and are determined by an average of the two linear responses spanning the plane of rotation of the pulse. 
Data shown in Fig.\ref{fig:mn2au_110_condTens_vs_hw}(b) 
reveal that  the aforementioned opposite sign between $J_{\perp\mathbf{N}}$ driven by light with $\boldsymbol{\varepsilon}\parallel\mathbf{N}$ and $\boldsymbol{\varepsilon}\perp\mathbf{N}$ (red and blue lines)  is robust with respect to the band filling for a range of almost 2\,eV around the true Fermi level.
At about 0.5\,eV below the true Fermi level the photocurrents for  two in-plane linear polarizations have identical magnitude and opposite sign $-$ this can be directly seen from the vanishing response to circularly polarized light rotating in the plane of the N\'eel vector (pink symbols), which vanishes there as a result of being an average over two  currents of opposite sign. Notably, the band filling behavior is manifestly different from that exhibited by the photocurrents at another frequency and band smearing, shown in Fig.\ref{fig:mn2au_3at_100vs010}, and we scrutinize the sensitivity of the photocurrents to disorder next.
%
\subsubsection{Effect of quasiparticle lifetime.}

In this section the dependence of the photocurrents on the quasiparticle lifetime when approaching the clean limit (i.e.~broadening parameter $\Gamma\to0$)  is analyzed. We consider positive and negative $\Gamma$ values to identify the magnetic origin of photocurrents. By analyzing the $\Gamma$-scaling of the photocurrents it is also possible to identify the role of different quantum geometric quantities\cite{watanabe2020nonlinear,holder2021mixed,kaplan2022unification,ahn2022riemannian}. General understanding of current photo-response suggests that three terms can be identified which differ in their scaling behavior with respect to $\Gamma$: a Drude term quadratic in the lifetime broadening, a term probing the Berry curvature dipole which is linear in the lifetime broadening, and a term probing the quantum metric dipole independent on the lifetime broadening\cite{watanabe2020nonlinear}. However, the Berry curvature itself and therefore the Berry curvature dipole term are expected to vanish in $\mathcal{PT}$-symmetric systems \cite{watanabe2020nonlinear} leaving only the Drude and the quantum metric dipole terms.

\begin{figure} 
    \centering
      \includegraphics[width=0.7\columnwidth]{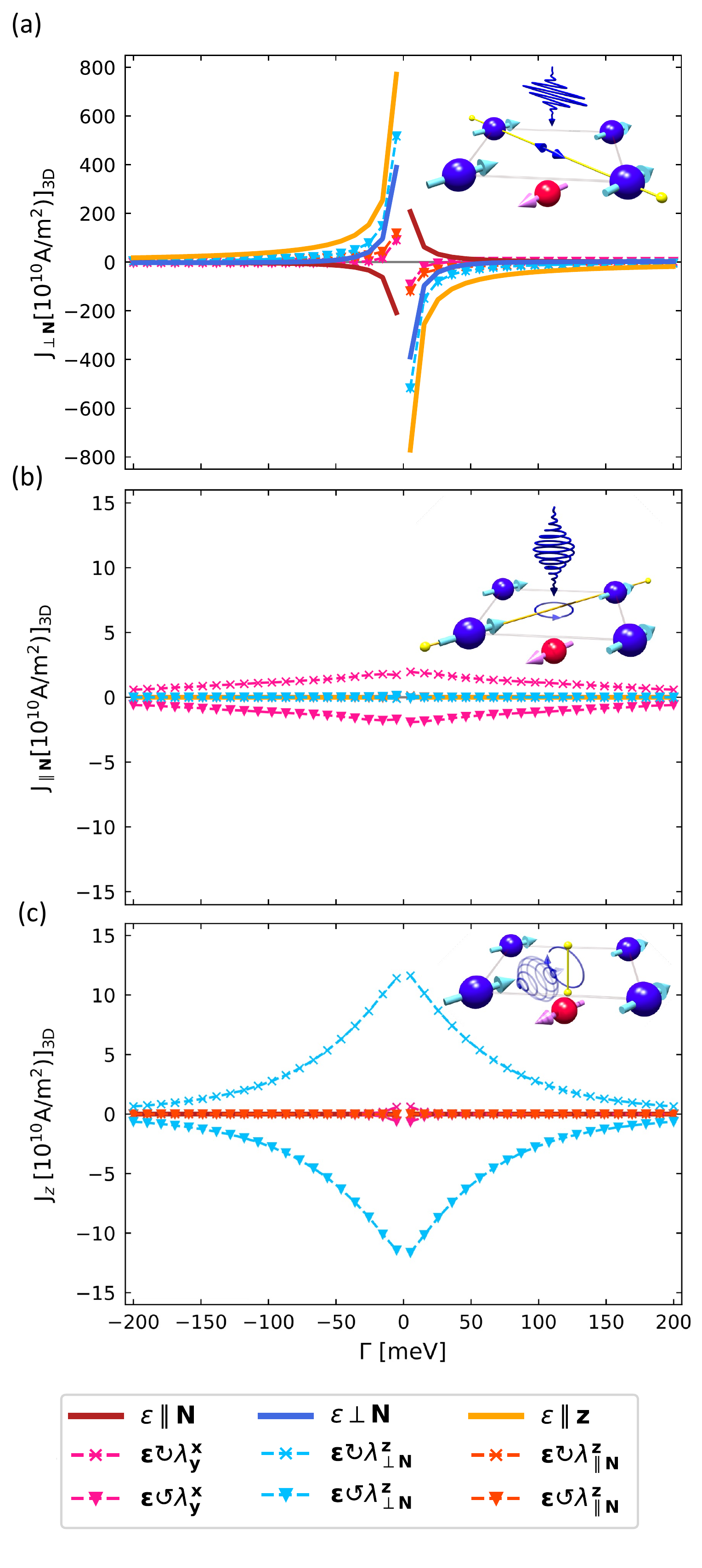}
    \caption[]{Dependence of the photocurrents on the broadening parameter $\Gamma$ at the true Fermi energy of Mn$_2$Au.  The dependence of (a) $J_{\perp\mathbf{N}}$, (b) $J_{\Vert\mathbf{N}}$ and (c) $J_z$ is shown for $\hbar\omega=3.00\,$eV for different types of laser pulses. 
    For more details see text.  
    }
    \label{fig:mn2au_110_epC_vs_posNegSMR} 
\end{figure}

We show in Fig.\ref{fig:mn2au_110_epC_vs_posNegSMR}(a-c) the results of our calculations for $J_{\perp\mathbf{N}}$, $J_{\Vert\mathbf{N}}$ and $J_z$ currents arising in response to different types of pulses of frequency $\hbar\omega=3.0$\,eV 
in a moderate range of positive and negative $\Gamma$ values up to 200 meV. 
We observe that at 25\,meV broadening $J_{\perp\mathbf{N}}$ is as large as $-150\times 10^{10}$\,A/m$^2$ in case of linearly $z$-polarized light [orange line in (a)], and a comparable current magnitude for the other  two cases of in-plane polarizations [red and blue lines in (a)] is visible. Sizeable responses to circularly polarized light $-$ which are helicity independent averages of the response for two linear polarizations  defining the plane of rotation $-$ are also visible in $J_{\perp\mathbf{N}}$. 
These contributions can be thus associated with the LPGE. On the other hand, the other two components of the photocurrent $-$ $J_{\parallel\mathbf{N}}$ and $J_z$ $-$ are an order of magnitude smaller over the entire range of $\Gamma$ but they are helicity-switchable. These photocurrents can thus be associated with the CPGE. 
While the LPGE currents are sizeable for all three shown directions of linear polarization, the CPGE responses are only present for a specific plane of rotation of light for a given direction of the resulting current and given N\'eel vector orientation. 

Notably, computed LPGE and CPGE responses are qualitatively different with respect to a reversal in the sign of $\Gamma$. This is indeed expected, as it is known that the LPGE and CPGE currents exhibit  different behavior under time reversal operation $\mathcal{T}$:
it can be shown that the LPGE is even under $\mathcal{T}$ for magnetization-independent,  and $\mathcal{T}$-odd for magnetization-dependent components of the conductivity tensor. However in $\mathcal{PT}$-symmetric materials the $\mathcal{P}$-symmetry is broken only by the magnetization, and not by the crystal structure itself, and the magnetization-independent part of LPGE is expected to vanish. Indeed such a magnetization dependent linear photogalvanic effect was predicted in 2D antiferromagnetic $\mathcal{PT}$-symmetric insulator CrI$_3$, and proposed for 3D $\mathcal{PT}$-symmetric AFMs such as Mn$_2$Au\cite{felser2019switchable}.  

The behavior of the computed currents under $\Gamma$-reversal can be now understood by realizing that switching the sign of the quasiparticle lifetime is equivalent to a reversal of the N\'eel vector. Indeed the linear photocurrent $J_{\perp\mathbf{N}}$ is odd with respect to the broadening, while the circular responses $J_{\parallel\mathbf{N}}$ and $J_{z}$ are even in $\Gamma$. In conclusion the LPGE photocurrents can be switched by a reversal of the Mn magnetic moments and, at certain laser frequencies, also by changing the orientation of the linearly polarized light. The CPGE photocurrents can only be switched by changing the light helicity and are one order of magnitude smaller as compared to the LPGE photocurrents. 
  \cite{wang2020electrically}

\section{Inverse Faraday Effect}\label{section4}

\begin{figure}[h]
    \centering
      \includegraphics[width=0.98\columnwidth]{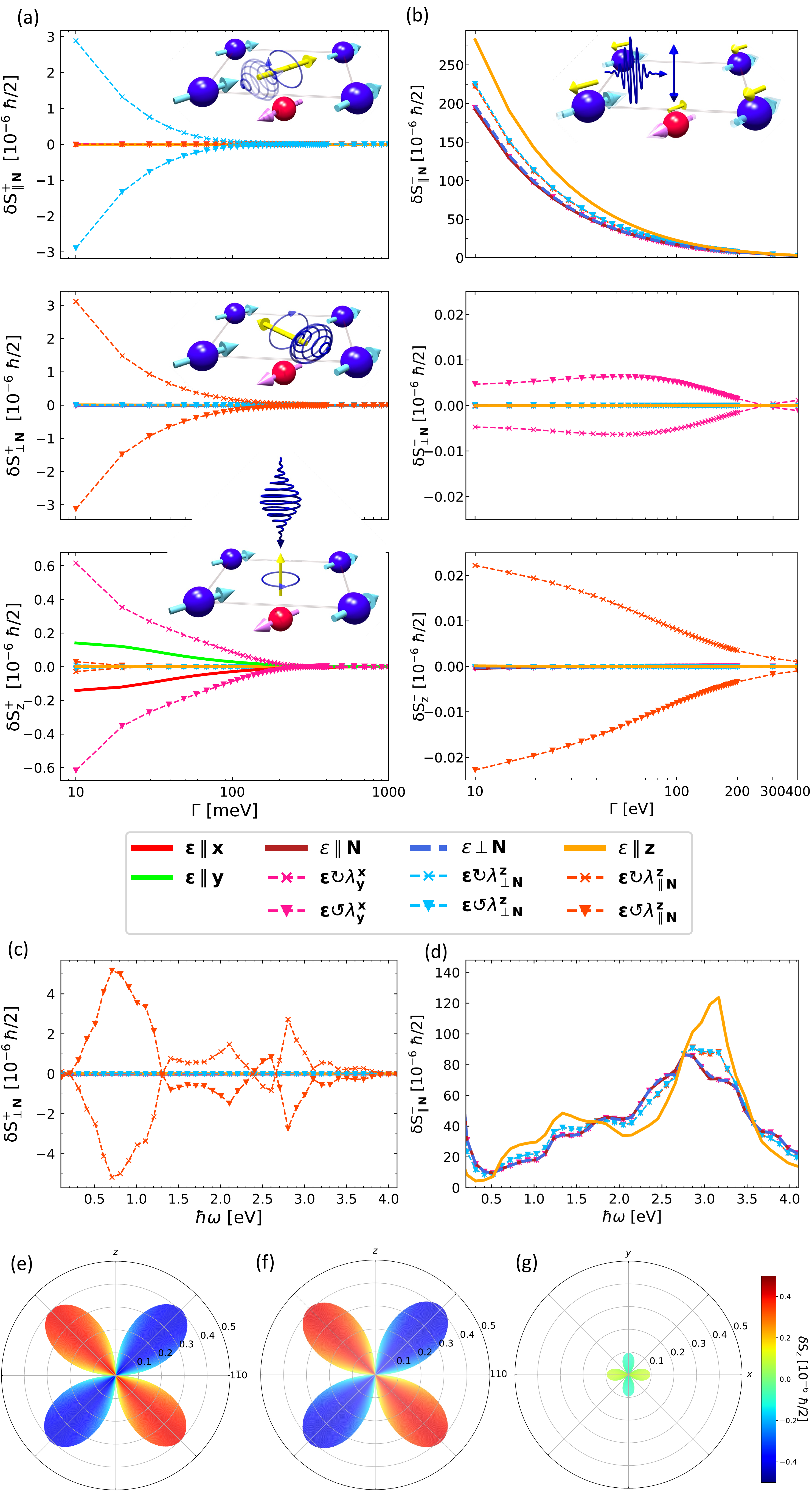}
    \caption[]{Inverse Faraday effect in Mn$_2$Au. (a,b) Shown is  the $\Gamma$ dependence of total (a) and staggered (b) spin density, $\delta S^+$ and $\delta S^-$ respecitively,  for a laser frequency of $\hbar\omega=3.00\,$eV.   In (c,d) the frequency dependence of the total (c) and staggered (d) laser induced spin, perpendicular and parallel to the N\'eel vector, respectively, is shown for $\Gamma=25$\,eV.
     In (e,f,g) the dependence of (e) parallel, $\delta S^+_{\Vert\mathbf{N}}$, (f) perpendicular, $\delta S^+_{\perp\mathbf{N}}$, and (g) $z$-component, $\delta S^+_z$, of the total spin density on the orientation of polarization vector of linearly polarized light is shown. In each case, the polarization is rotated in the same plane in which the circularly polarized light rotates in (a).   
    }
    \label{fig:mn2au_spD_vsSmr_hw3000}
\end{figure}

In this section, we discuss the properties of the light-induced spin polarization also known as the {\it inverse Faraday effect} (IFE) $-$ which is believed to play a crucial role in optical switching and optical dynamics of antiferromagnets\cite{ziel1965IFE,vahaplar2009ultrafast,oppeneer2016ab,freimuth2016torques,dannegger2021ultrafast,oppeneer2021ultrafast} $-$ in Mn$_2$Au. 
We compute and analyze the overall induced spin moment in the unit cell, $\delta S^{+}$, defined as the sum over all induced spin moments in the unit cell, as well as the {\it staggered} component of the induced spin, defined as the difference between the moments on different Mn sublattices: $\delta S^{-}=\frac{1}{2}[\delta S({\rm Mn_A})-\delta S({\rm Mn_B})]$. 
The results of our calculations of $\delta S^{+}$ and $\delta S^{-}$ as a function of $\Gamma$ (at $\hbar\omega=3.0$\,eV) and frequency (at $\Gamma=25$\,meV) are presented in Fig.\ref{fig:mn2au_spD_vsSmr_hw3000}.
Similarly to previously discussed photocurrents the induced spin is projected onto the $z$-, $\parallel\mathbf{N}$- and $\perp\mathbf{N}$-directions.

We observe that the integrated induced spin,  Fig.\ref{fig:mn2au_spD_vsSmr_hw3000}(a), is most sizeable for the two in plane components $\delta S^+_{\Vert\mathbf{N}}$ and  $\delta S^+_{\perp\mathbf{N}}$, which exhibit a magnitude of the order of $1 \times 10^{-6}\,\hbar/2$ at 25\,meV broadening at the frequency of $\hbar\omega=3.00$\,eV. All components are finite only when the direction of the induced spin density is along the normal of the plane of rotation of circularly polarized light, and they can be switched by the sense of light helicity, as expected from the IFE\cite{ziel1965IFE}. Each component, however, shows a unique frequency dependence: as Fig.\ref{fig:mn2au_spD_vsSmr_hw3000}(c) shows, the frequency dependence of the perpendicular in-plane spin density $\delta S^+_{\perp\mathbf{N}}$ can reach up to $5 \times 10^{-6}\,\hbar/2$ at the frequency of 800\,meV. The in-plane component which is parallel to the N\'eel vector (not shown) on the other hand does not exceed a value of $\sim 1.5 \times 10^{-6}\,\hbar/2$ throughout the entire frequency range.   

The out-of-plane component of the induced spin $\delta S^+_z$ is significantly smaller than the other two components: 
at 25\,meV of broadening $\delta S_z^+\approx 0.35\times 10^{-6}\,\hbar/2$ which is more than two times smaller than the in-plane components. However, it shows the slowest decay towards larger broadening values, and it is comparable in magnitude to the in-plane components for $\Gamma$ above 100\,meV. In the range of $\Gamma$ from 20\,meV up to almost $200$\,meV the $z$-component shows an almost perfect linear decrease when plotted versus $\Gamma$ on a logarithmic scale indicating a $\sim \log(\Gamma)$ behavior, whereas the in-plane components show a much quicker decay in this region. 

Recently, we predicted that a laser-induced torque in Mn$_2$Au can be generated not only by circular but also by linearly polarized light.\cite{freimuth2021torques}
Our caculations of the IFE are in complete agreement with the results of the latter work in terms of symmetry of linear $\delta S^+$ and $\delta S^-$.
For example, we can see in Fig.\ref{fig:mn2au_spD_vsSmr_hw3000}(a) that a non-vanishing  $\delta S^{+}_z$ can arise in response to light polarized linearly along $x$- and $y$-direction, but not when the light polarization 
is along the $\perp\mathbf{N}$ and $\parallel\mathbf{N}$ directions. 
The overall dependence of $\delta S^{+}_z$, with $\hbar\omega=3.00$\,eV and $\Gamma=25$\,meV, on the orientation of linearly polarized light within the $xy$-plane is shown  in Fig.\ref{fig:mn2au_spD_vsSmr_hw3000}(g). In analogy to $\delta S^{+}_z$, the components $\delta S^+_{\parallel\mathbf{N}}$ and $\delta S^+_{\perp\mathbf{N}}$ can also be driven by linearly polarized light: if the light polarization is confined to the plane whose normal is aligned with the induced spin density (the same plane in which a finite response to circularly polarized light occurs). Figures~\ref{fig:mn2au_spD_vsSmr_hw3000}(e,f) display the dependence of $\delta S^+_{\parallel\mathbf{N}}$ and $\delta S^+_{\perp\mathbf{N}}$ on the orientation on linear polarized light in the corresponding planes.

While the magnitude of the in-plane responses is overall significantly larger than that of $\delta S^+_z$, the dependence of the sign of the spin polarization on the light polarization  and a non-trivial angular dependence of the signal suggests that a very complex spin response in Mn$_2$Au can be driven by the ``off-axis'' linearly polarized pulses.
While the uniform response $\delta S^+$ is effectively equivalent to an effect of canting or formation of weak ferromagnetism, the staggered component of IFE, $\delta S^-$, can give rise to a N\'eel-type torque on the staggered magnetization\cite{wadley2015,zelezny2015}, or result in a staggered modulation of the local moments. As for the latter, while the N\'eel type torques are being currently intensively explored with respect to electrical and optical switching of antiferromangetic order,\cite{olejnik2018terahertz,bodnar2018writing,mokrousov2021roadmap,wunderlich2022ultrashort,oppeneer2023neel} recently a large staggered IFE was reported in AFM CrPt \cite{oppeneer2021ultrafast}, were a direct quenching of the magnetic moments on the Cr sublattices was achieved by  circularly polarized light. The reported quenching was identified as the origin of ultrafast optical switching of the N\'eel vector in CrPt \cite{oppeneer2021ultrafast}, highlighting the importance of staggered IFE in AFM materials.
Notably, the properties of $\delta S^-$, that we compute and present in Fig.\ref{fig:mn2au_spD_vsSmr_hw3000}(b), are quite different from the properties of $\delta S^+$ in many respects.  

First, the staggered IFE components perpendicular to the N\'eel vector, $\delta S^-_{\perp\mathbf{N}}$ and  $\delta S^-_{z}$,  are one to two orders of magnitude smaller than the
corresponding components of $\delta S^+$, and they exhibit a qualitatively different behavior with disorder. Second,
the staggered response parallel to the N\'eel vector, $\delta S^-_{\parallel\mathbf{N}}$, is two orders of magnitude larger then $\delta S^+$, reaching  as much as  $10^{-4}\,\hbar/2$ at room temperature ($\Gamma=25$\,meV) in value. Moreover, the staggered response is finite not only for circularly polarized light but also for linearly polarized light, with circular responses being helicity-independent. Thus, among all discussed components of $\delta S^-$ and $\delta S^+$, $\delta S^-_{\parallel\mathbf{N}}$ behaves like a LPGE effect while all the other components exhibit a CPGE-like behavior. 
Remarkably, also the frequency dependence of $\delta S^-_{\parallel\mathbf{N}}$ is very distinct: as we see in  Fig.\ref{fig:mn2au_spD_vsSmr_hw3000}(d), 
in the entire frequency range the staggered response has the same amplitude for light linearly or circularly polarized within the $xy$-plane. For the latter, the largest response of $85\times 10^{-6}\hbar/2$ occurs at $\hbar\omega=2.8$\,eV. For linear polarization along $z$ an even larger peak of $125\times 10^{-6}\hbar/2$ is visible at $\hbar\omega=3.1$\,eV. Note, that also the sign of $\delta S^-_{\parallel\mathbf{N}}$ is preserved for the entire frequency band, which is in stark contrast to other components. Overall, the complexity of laser-induced flavors of spin polarization in Mn$_2$Au suggests that the physics of laser-induced spin currents in this material, which we consider next, can be also very rich.

\begin{figure}
    \centering
    \includegraphics[width=1.0\columnwidth]{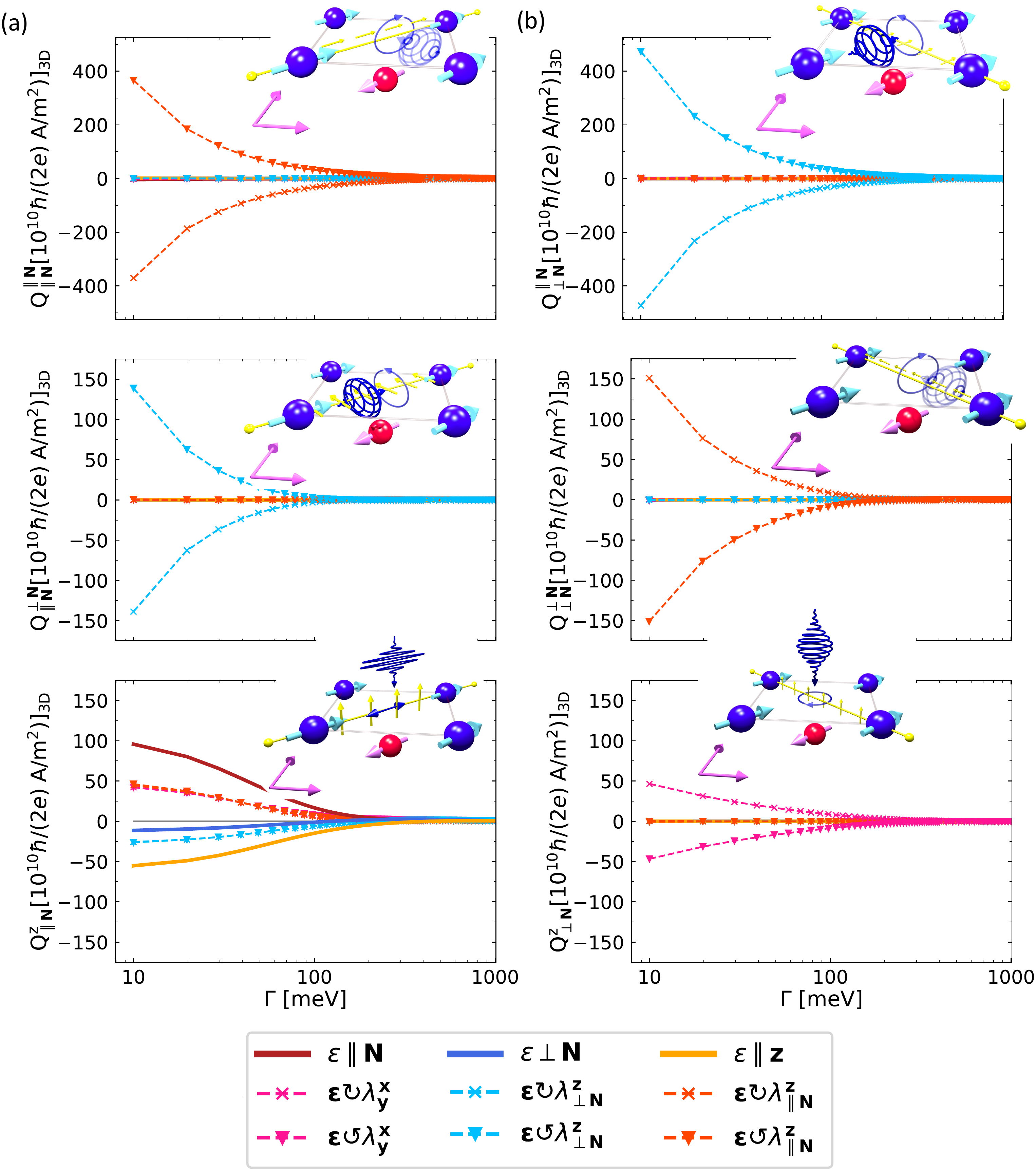}
    \caption[Laser induced spin currents in Mn$_2$Au]{Laser-induced spin currents as a function of  broadening $\Gamma$ for a frequency of $\hbar\omega=3.00\,$eV. Left column, (a), shows the spin currents flowing parallel to the N\'eel vector, the right column, (b), perpendicular to the N\'eel vector for different directions of spin polarization as indicated by yellow arrows. In our nomenclature, $Q^{\mathbf{B}}_{\mathbf{A}}$ denotes a spin current propagating along direction $\mathbf{A}$ having spin polarization along direction $\mathbf{B}$.}
    \label{fig:mn2au_spC_vsSmr_hw300}
\end{figure}

\section{Laser induced Spin currents}\label{section5}


In solids, interaction of light with matter mediated by spin-orbit interaction suggests an exciting possibility of generating pure spin currents with light.
Historically, nonmagnetic semiconductors were first found to host a spin Hall effect\cite{wunderlich2005experimental}, which was  suggested to generate spin currents  under radiation with circularly polarized light \cite{sih2006generating}. Later on, an all-optical generation of spin currents via a transfer of optical angular momentum into a spin-orbit-coupled semiconductor was reported \cite{watzel2018all}. Recently, an optical generation of colossal spin currents which are second order in the field has been predicted theoretically for ferromagnets \cite{freimuth_2021}. Moreover, $\mathcal{PT}$-symmetric collinear AFMs, such as discussed here  Mn$_2$Au, were proposed as efficient sources of spin currents \cite{hayami2022nonlinear}, and a large spin circular photogalvanic effect was predicted for $\mathcal{PT}$-symmetric antiferomagnetic insulators \cite{fei2021symmetry}. It was also shown that in non-centrosymmetric systems which still hold a mirror symmetry, pure bulk spin currents can be generated even by means of linearly polarized light \cite{xu2021pure}.%
%

\begin{figure}
    \centering
    \includegraphics[width=0.72\columnwidth]{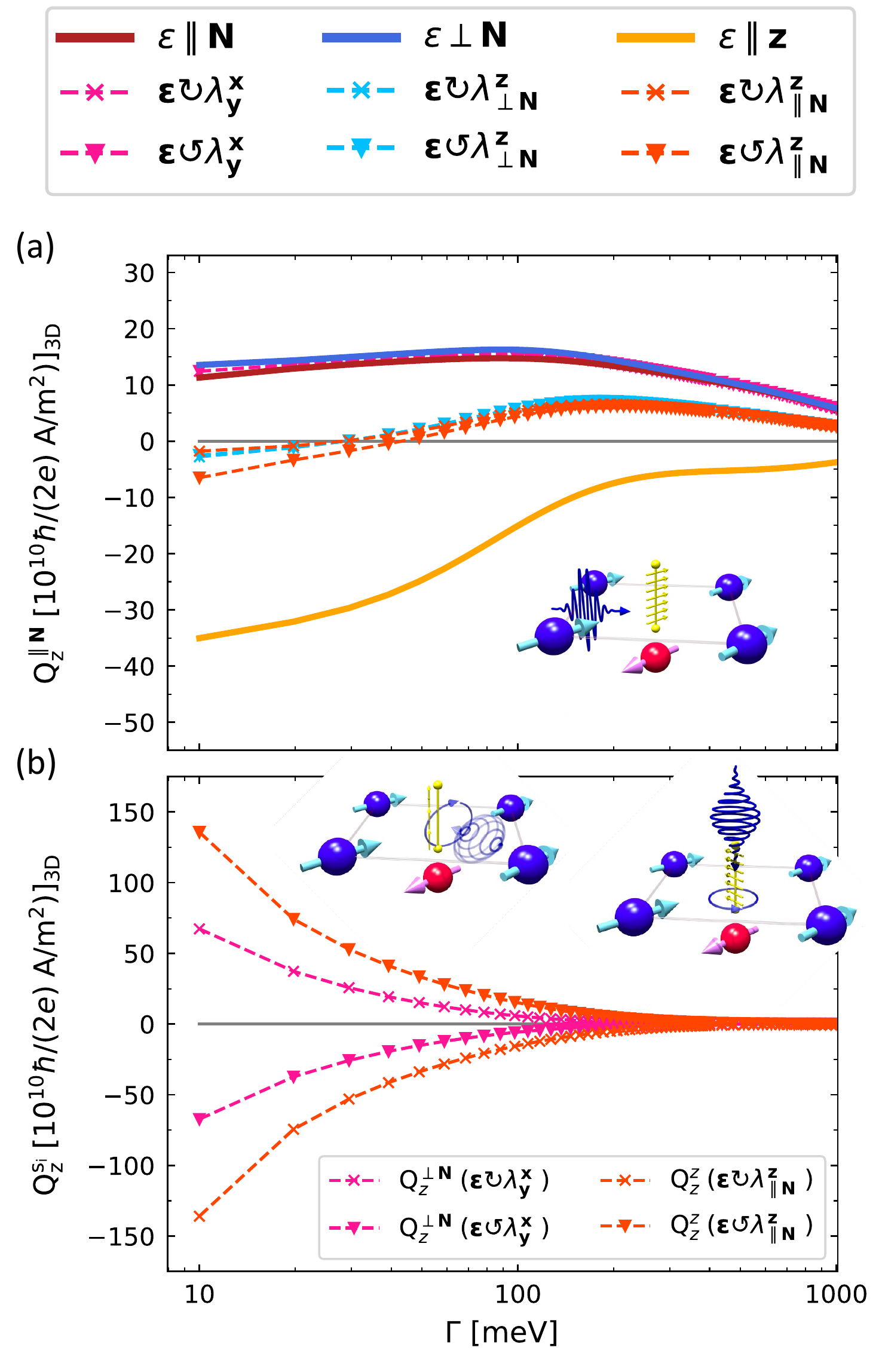}
    \caption[Laser induced spin currents in Mn$_2$Au]{Laser-induced spin currents as a function of  broadening $\Gamma$ for a frequency of $\hbar\omega=3.00\,$eV. (a) Spin currents flowing along $z$ with spin polarization along the N\'eel vector. (b) Spin currents flowing along $z$ for spin polarization along $z$ and perpendicular to the N\'eel vector.  }
    \label{fig:mn2au_spC_z_vsSmr_hw300}
\end{figure}

In this section, we investigate laser-induced spin-polarized photocurrents, or {\it spin photocurrents}, arising in response to linearly and circularly polarized light, in Mn$_2$Au. We compute the magnitude of a spin current $Q^{\mathbf{B}}_{\mathbf{A}}$, propagating along direction $\mathbf{A}$ with spin polarization along direction $\mathbf{B}$, as a function of $\Gamma$ at $\hbar\omega=3.0$\,eV, presenting the results in Fig.\ref{fig:mn2au_spC_vsSmr_hw300} and Fig.\ref{fig:mn2au_spC_z_vsSmr_hw300}.
First, we discuss the spin currents propagating  in the $xy$-plane, shown in Fig.\ref{fig:mn2au_spC_vsSmr_hw300}.
In the latter case, the most sizeable spin photocurrents are $Q^{\parallel\mathbf{N}}_{\parallel\mathbf{N}}$ and  $Q^{\parallel\mathbf{N}}_{\perp\mathbf{N}}$ (Fig.\ref{fig:mn2au_spC_vsSmr_hw300}, upper panel) achieved by circularly polarized light 
with in-plane direction of the  polarization plane normal, 
with amplitudes reaching as much as $\sim150\times10^{10}$\,$\hbar/(2\mathrm{e})\,\mathrm{A/m^2}$ at room temperature  ($\Gamma=25\,$meV). Both components can be switched in sign by light helicity. 
The components with the spin polarization perpendicular to the N\'eel vector,
$Q^{\perp\mathbf{N}}_{\parallel\mathbf{N}}$ and  $Q^{\perp\mathbf{N}}_{\perp\mathbf{N}}$ (Fig.\ref{fig:mn2au_spC_vsSmr_hw300}, middle panel),
are somewhat smaller in magnitude but still significant. From the plots, an apparent symmetry is visible: at a given direction of spin polarization, along or perpendicular to $\mathbf{N}$, the direction of the current in-plane can be rotated by changing the in-plane direction of polarization plane normal.



The spin current flowing parallel to the N\'eel vector with spins pointing along the $z$-direction, $Q^{z}_{\parallel\mathbf{N}}$, shown in the lower panel of Fig.\ref{fig:mn2au_spC_vsSmr_hw300}, is the only in-plane component responding to linearly polarized light, with amplitudes reaching $75\times10^{10}$\,$\hbar/(2\mathrm{e})\,\mathrm{A/m^2}$ at $\Gamma=25\,$meV. Here, the corresponding circular spin currents are also sizeable but are helicity-independent, and are given by the averages of the linear polarized responses $-$ this marks $Q^{z}_{\parallel\mathbf{N}}$ as a manifestation of LPGE. In analogy to the case of charge photocurrents shown in Fig.\ref{fig:mn2au_110_epC_vs_posNegSMR}, LPGE-like $Q^{z}_{\parallel\mathbf{N}}$ decays much slower with $\Gamma$ then the rest of spin currents, which are CPGE-like. 
Again in analogy to charge photocurrents,  the spin polarized current $Q^{z}_{\parallel\mathbf{N}}$ changes sign when rotating the in-plane linear polarization by 90$^\circ$ in the $xy$-plane (red and blue solid lines) or into $z$-axis (orange solid line). 
A given linearly polarized pulse thus not only induces a charge current flowing perpendicular to $\mathbf{N}$, $J_{\perp\mathbf{N}}$, but also a spin-polarized current $Q^{z}_{\parallel\mathbf{N}}$, which is perpendicular to $J_{\perp\mathbf{N}}$. This can be interpreted as an {\it optical analogon of the spin Hall effect}, as discussed in the next subsection (see also Fig.\ref{fig:mn2au_110_sH_coeff_cartoon}).

Finally, in Fig.\ref{fig:mn2au_spC_z_vsSmr_hw300} we display the spin photocurrents flowing out of the plane along the $z$-direction for different orientations of the spin polarization. Two types of spin currents can be distinguished here: the LPGE-like spin current $Q^{\Vert\mathbf{N}}_{z}$, shown in  Fig.\ref{fig:mn2au_spC_z_vsSmr_hw300}(a), and CPGE-like helicity-dependent spin currents, with spin polarization perpendicular to $\mathbf{N}$, shown in Fig.\ref{fig:mn2au_spC_z_vsSmr_hw300}(b). Among the two flavors of spin currents, in accord to our previous observations, LPGE-currents are subdominant and exhibit a non-trivial dependence on broadening: the majority of circular currents, being an average over linear currents of opposite sign and non-trivial $\Gamma$-dependence, change their sign at the value of broadening of about 30\,meV.
By generalizing our observations, we can conclude that spin currents with spins aligned with the direction of propagation, i.e. $Q^{a}_{a}$ with $a\in \{\parallel\mathbf{N},\perp\mathbf{N},z\}$, only respond to circularly polarized light rotating in the plane spanned by the N\'eel-vector and the $z$-axis.  
On the other hand,
spin currents of the type $Q^{b}_{a}$, for $a\ne b \in \{\parallel\mathbf{N},\perp\mathbf{N},z\}$, are always driven by the same type of pulse as $Q^{a}_{b}$ spin currents.

We recall that LPGE charge photocurrents, presented in Fig.\ref{fig:mn2au_110_epC_vs_posNegSMR}, were found to be odd under sign reversal of the broadening, while CPGE charge photocurrents showed an even behavior, identifying both as purely magnetic photocurrents. In the Keldysh formalism, the difference between the charge and spin photocurrent comes from replacing one of velocity operators $v_i$ 
with the spin-velocity operator $\{v_i,\tau_s\}$, where $\tau_s$ is one of the Pauli matrices. 
Since $\tau_s$ is odd under $\mathcal{T}$, the behavior of spin photocurrents with respect to the reversal in $\Gamma$ is opposite to that of charge photocurrents. 
Magnetic LPGE-like spin photocurrents therefore are expected to be even under sign reversal of $\Gamma$ and magnetic CPGE-like spin-photocurrents odd. We  checked the behavior of the spin currents, presented in Fig.\ref{fig:mn2au_spC_vsSmr_hw300} and Fig.\ref{fig:mn2au_spC_z_vsSmr_hw300}, with respect  to a sign reversal of the broadening, and found all, LPGE- and CPGE-like, responses to be purely magnetic. 
\subsection{Photospin Hall Effect in Mn$_2$Au}\label{section5a}
At this point, let us summarize our findings for a simple case of light, linearly polarized along $\mathbf{N}$, $\mathbf{N}||[110]$. In this case, the only type of a photoinduced charge current is $J_{{\perp}\mathbf{N}}$, the only non-vanishing photo-induced spin polarization is $\delta S^-_{\Vert \mathbf{N}}$, and two types of pure spin current: $Q^{\Vert\mathbf{N}}_{z}$ and $Q_{\Vert\mathbf{N}}^{z}$. Both of these spin currents nominally satisfy the spin Hall effect geometry: i.e. the direction of spin current propagation and its spin polarization are orthogonal to each other, and to the direction of $J_{{\perp}\mathbf{N}}$. Of these two, we will consider below only $Q_{\Vert\mathbf{N}}^{z}$, since it is larger in magnitude than the former (which is nominally a manifestation of anisotropic spin Hall effect in SHE language\cite{freimuth2010anisotropic}), while their properties, discussed below, are similar. It therefore seems insightful to scrutinize a possible relation between the three LPGE-like types of quantities $J_{{\perp}\mathbf{N}}$,  $Q_{\Vert\mathbf{N}}^{z}$ and $\delta S^-_{\Vert \mathbf{N}}$.


We first compare the behavior of $J_{{\perp}\mathbf{N}}$ and $Q_{\Vert\mathbf{N}}^{z}$ by plotting in Fig.\ref{fig:mn2au_sge_vs_hw}(a,b) their frequency dependence for three flavors of the linearly polarized light.
What we realize immediately is that the considered charge and spin currents exhibit some qualitative similarities with respect to $\omega$.  
Both charge and spin photocurrents are enhanced in the region of smaller frequencies, where the responses are comparable in amplitude for two in-plane polarizations (red and blue curve), changing their sign at a frequency of $\hbar\omega=1.00\,$eV. In the frequency band of $\hbar\omega$ from 2\,eV to $3.5\,$eV a strong oscillatory behavior is visible for light linearly polarized in the $xy$-plane. At $\hbar\omega=2.8\,$eV the spin photocurrent is maximized with a strong peak for light linearly polarized along $\mathbf{N}$ (red curve) and a negative peak for light polarized along  $z$ (golden curve). A corresponding peak for $z$-polarization is also visible at this frequency for the charge current.

\begin{figure}[t] 
    \centering
    \includegraphics[width=0.75\columnwidth]{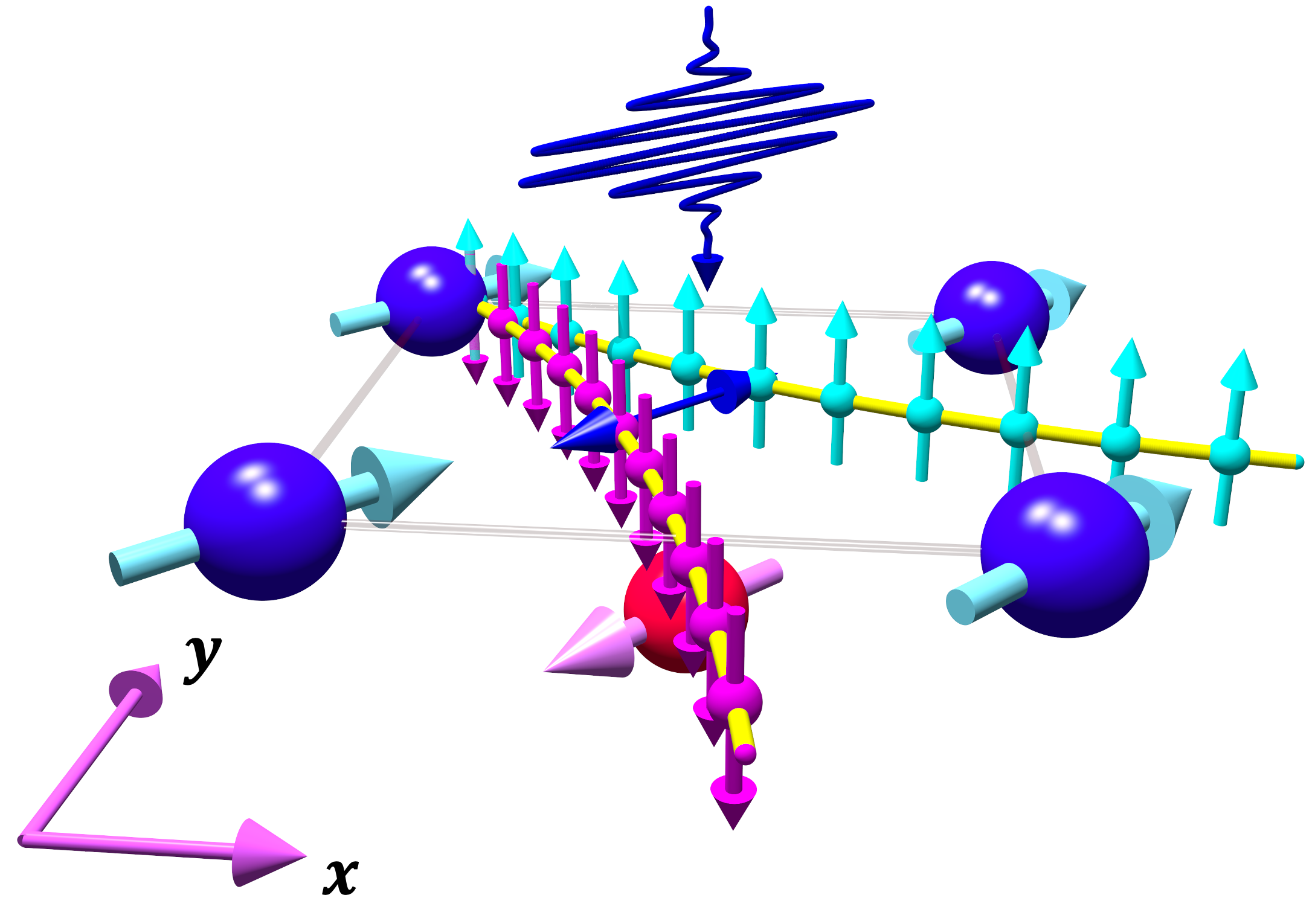}
    \caption[PSHA in Mn$_2$Au]{Sketch of the  photospin Hall effect in Mn$_2$Au. In Mn$_2$Au a linearly polarized light (shown with a wavy arrow, the polarization direction shown with a blue two-sided arrow in the center of the plot), generates an in-plane charge photocurrent perpendicular to N\'eel vector $\mathbf{N}$ (given by the direction of the arrow on the ``blue" sublattice of Mn atoms). Via the photospin Hall effect, the charge current generates a transverse spin current with spin polarization out of the plane, propagating along $\mathbf{N}$. The effect of the spin-Hall-like spin-dependent deflection of electrons (small balls), whose spin is depicted with vertical blue and cyan little arrows, is shown by an effective curving of the electron trajectories.}
    \label{fig:mn2au_110_sH_coeff_cartoon}
\end{figure}

Given the proper symmetry of the effect, as discussed above, it seems rewarding to interpret the relation between $J_{{\perp}\mathbf{N}}$ and $Q_{\Vert\mathbf{N}}^{z}$ in terms of a flavor of spin Hall effect, see 
Fig.~\ref{fig:mn2au_110_sH_coeff_cartoon}.
We suggest to refer to this effect as the {\it photospin Hall effect}, which can be characterized in terms of a {\it photospin Hall angle} $\Theta_{\rm PSH}$, PSHA, defined as the normalized ratio
between the charge and spin photocurrents, transverse to each other: 
%
%
%
\begin{equation}
    \Theta_{\mathrm{PSH}}    =    \arctan{\frac{Q^{z}_{\parallel \mathbf{N}}}{J_{\perp\mathbf{N}}}}.
    \label{eq:spinHall_angle}
\end{equation}
The practical meaning of the photospin Hall angle lies in providing an ability to predict the value of the generated transverse spin photocurrent, given an experimentally measured value of the charge photocurrent.
Fig.\ref{fig:mn2au_110_sH_coeff}(c,d) displays the computed PSHA in Mn$_2$Au for various flavors of linearly polarized light plotted as a function of band filling (for $\hbar\omega=3.0$\,eV) and laser frequency (at the true Fermi energy), computed at $\Gamma=25$\,meV.  Generally, we observe that around the true Fermi energy of Mn$_2$Au the PSHA is relatively constant, acquiring a largest value for light polarized along $\mathbf{N}$. 
\begin{figure}[t] 
    \centering
\includegraphics[width=1.00\columnwidth]{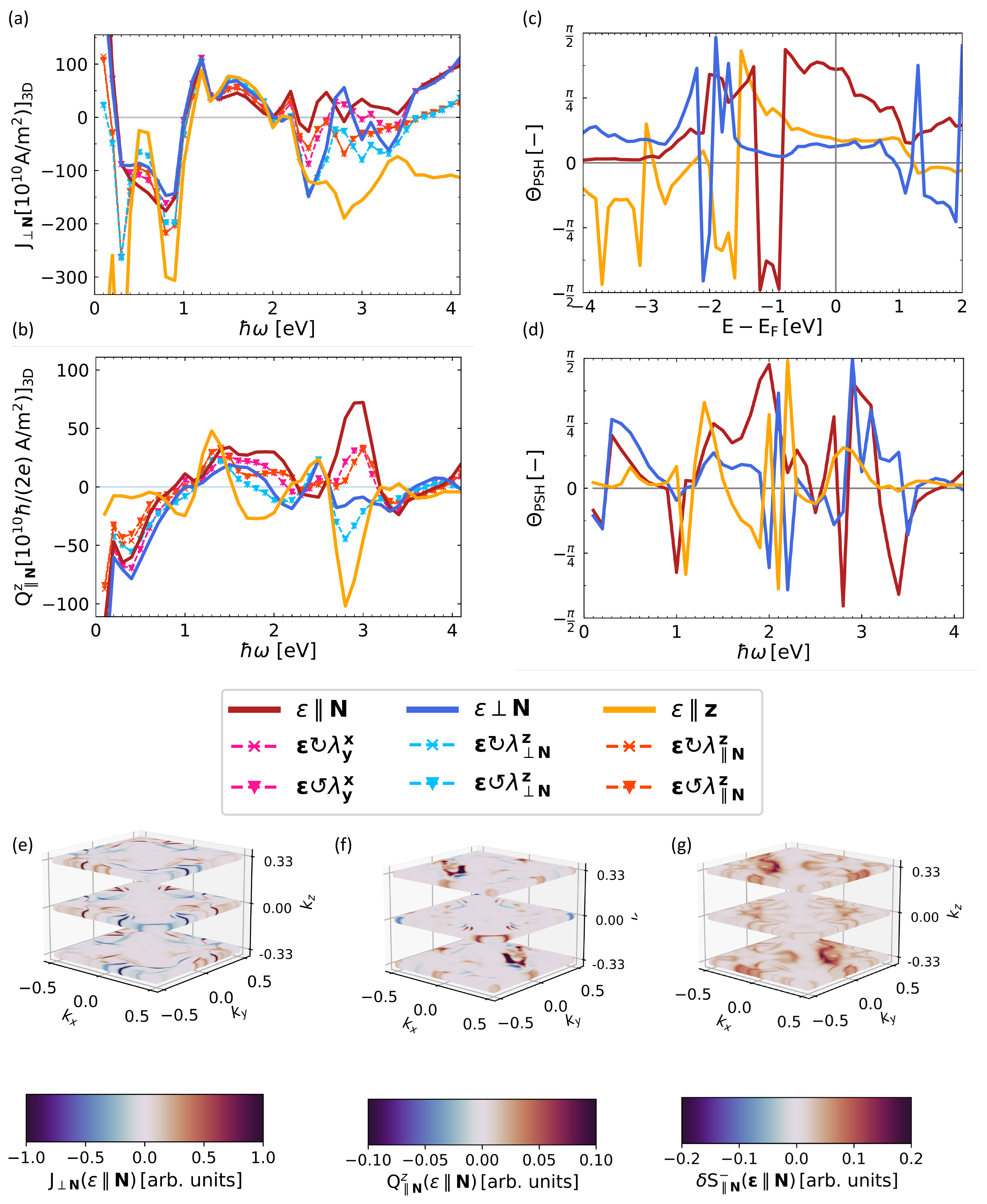}
    \caption[SGE in Mn$_2$Au]{Photospin Hall effect in Mn$_2$Au. (a-b) The frequency dependence of charge current $J_{\perp\mathbf{N}}$ (a) and spin current $Q_{\Vert\mathbf{N}}^{z}$ (b), for various types of incoming light pulse. In (a), the computed circular currents are averages of corresponding linear currents. In (c-d) the photospin Hall angle $\Theta_{\rm PSH}$, as defined in the text, is plotted versus the band filling at $\hbar\omega=3.0$\,eV (c), and frequency (d). (e-g) Brillouin zone distribution of $J_{\perp\mathbf{N}}$ (e),
    $Q_{\Vert\mathbf{N}}^{z}$ (f), and $\delta S^-_{\Vert \mathbf{N}}$ (g), at the frequency of
    $\hbar\omega=3.00\,$eV. For all calcualtions, the broadening $\Gamma$ was fixed to 25\,meV.}
    \label{fig:mn2au_110_sH_coeff}
    \label{fig:mn2au_sge_vs_hw}
\end{figure}
%
%
%
From the frequency dependence of the PSHA we learn about the rapidly oscillatory behavior of this quantity, suggesting that the ratio between the charge and spin currents can be tuned by the frequency not only in magnitude but also in sign.

To get a better feeling for the microscopic origin of the photospin Hall effect, we look at the reciprocal space distribution of considered charge and spin currents at $\hbar\omega=3.00\,$eV in Fig.\ref{fig:mn2au_110_sH_coeff}(e,f).
The charge photocurrent $J_{\perp\mathbf{N}}$, shown in Fig.\ref{fig:mn2au_110_sH_coeff}(e), exhibits competing regions of positive and negative sign in reciprocal space, which yield a net effect when integrated over the entire BZ. 
Shown in Fig.\ref{fig:mn2au_110_sH_coeff}(f) is the distribution of  $Q^{z}_{\parallel\mathbf{N}}$. Clearly, when compared to the charge photocurrent, the spin photocurrent displays a qualitatively different behavior in the BZ. Instead of competing regions of positive and negative sign the spin photocurrent is mostly driven by two symmetry-related hotspots in $+k_z$- and $-k_z$-layers which both contribute with the same sign but are an order of magnitude smaller as compared to the hotspots of charge photocurrent, which are also positioned closer to the BZ boundary. 



Let us now take a look at laser-induced spin polarization $\delta S^-_{\Vert \mathbf{N}}$. First of all, we find that the BZ distribution of this quantity, presented in Fig.\ref{fig:mn2au_110_sH_coeff}(g), 
is qualitatively very similar to that of the spin current, 
especially with respect to the position and relative sign of major contributing hotspots, which might be related to the spin nature of both effects.
Second, we also find some similarities between $\delta S^-_{\Vert \mathbf{N}}$ and $Q^{z}_{\parallel \mathbf{N}}$ in their dependence on laser frequency:
the spin photocurrent, Fig.\ref{fig:mn2au_sge_vs_hw}(f), exhibits local extrema at the frequency around $\hbar\omega=2.8\,$eV, which is also the case for photo-induced spin density, presented in Fig.\ref{fig:mn2au_spD_vsSmr_hw3000}(d).
In addition, for linearly polarized along $z$ light all three quantities $-$ the charge and spin photocurrents as well as nonequilibrium spin density $-$ display  a peak at $\hbar\omega=1.2$\,eV, which can be attributed to a large number of states available around that energy, see Fig.\ref{fig:mn2au_struct_comp}(c,d). Overall, we thus find some level of correlation between the three quantities, which is also expected in the conventional d.c. physics of the linear in the field Rashba-Edelstein-like phenomena.\cite{shen2014microscopic,freimuth2017pumping}

\section{Effect of canting}\label{section6}
Antiferromagnets are naturally prone to the effect of canting away from collinear configuration, which can be characterized by the so-called vector chirality proportional to the vector product between the spins on two sublattices. Recently,  we have shown that in bipartite canted antiferromagnets the behavior of the anomalous Hall effect (AHE) can be systematically categorized based on the  crystal\cite{smejkal2020crystal,feng2020observation} and chiral flavors of the AHE which are even and odd in vector chirality, respectively\cite{kipp2021chiral}. This separation has been also generalized to the domain of magneto-optical phenomena, and was shown to be crucial in understanding the temperature and magnetic field dependence of the AHE in antiferromagnets\cite{lux2022topological}. In AFMs, the chiral Hall effect directly probes the sense of chirality of the magnetic moments, as it was also shown for the case of  Mn$_2$Au.\cite{kipp2021chiral} In this section, we introduce and study the chiral flavors of the photocurrents and inverse Faraday effect in Mn$_2$Au.
 \subsection{Chiral photocurrents}\label{section6a}
\begin{figure} 
    \centering
    \includegraphics[width=0.95\columnwidth]{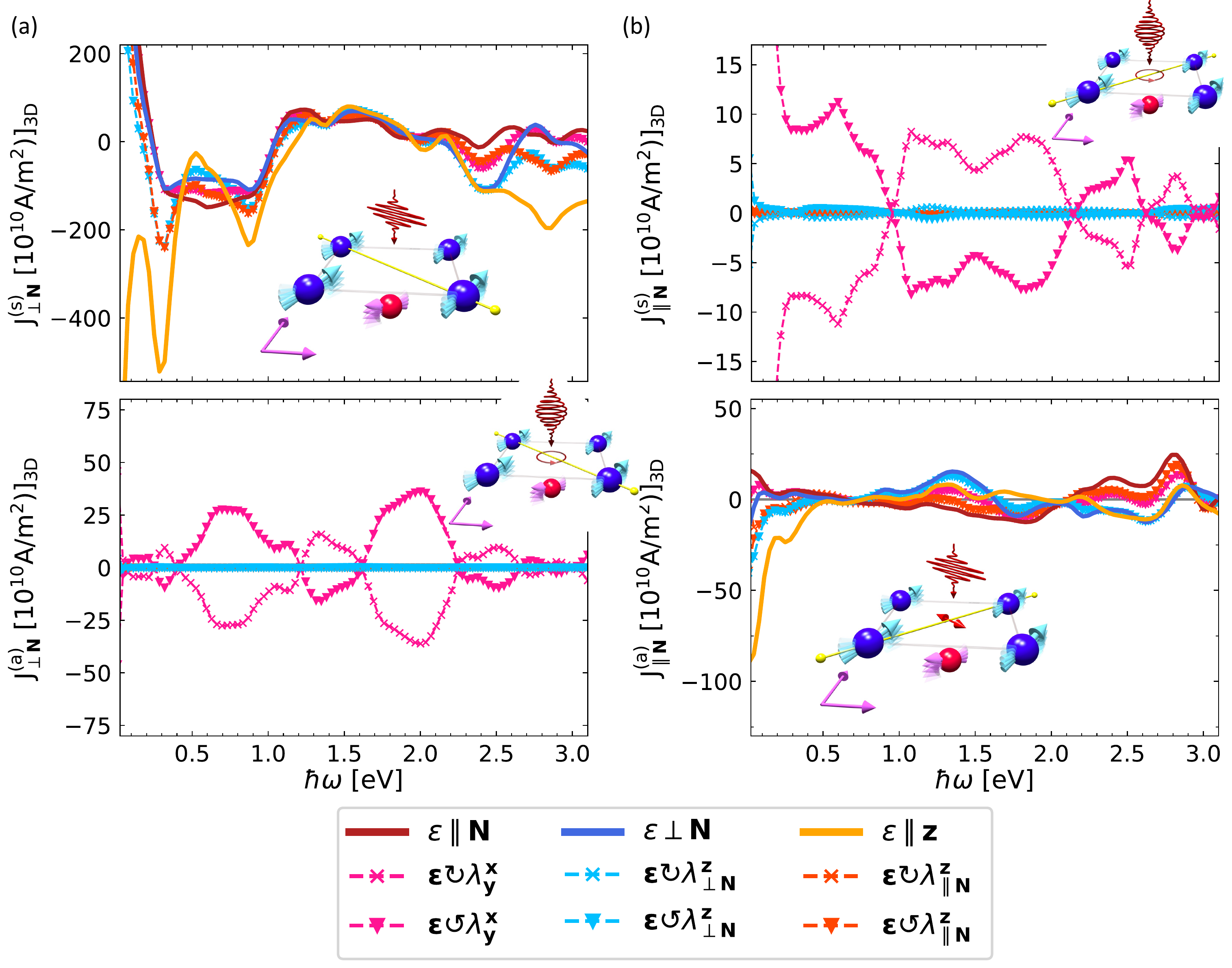}
    \caption[]{Symmetric (upper panel) and antisymmetric (lower panel) charge photocurrents at 200\,meV canting along $z$ which are (a) perpendicular, and (b) parallel to the N\'eel vector, as a function of the laser frequency for the broadening of 25\,meV.
    }
    \label{fig:mn2au_epC_para_zCant_Asymm}
\end{figure}

Similarly to the case of the anomalous Hall effect,\cite{kipp2021chiral} we introduce the {\it crystal} and {\it chiral photocurrents} as  symmetric $J^{(s)}$ and antisymmetric $J^{(a)}$ parts of the photocurrents upon switching the sign of chirality as given by the canting angle $\theta$ away from the $\parallel\mathbf{N}$-direction:
\begin{equation}
	J^{(s/a)}	=	\frac{J( + \theta ) \pm J(- \theta)}{2},
	\label{eq:symmAsymm_epC}
\end{equation}
where $J( \pm \theta )$ is the photocurrent evaluated in the system with spins canted by an angle $ \pm \theta$ towards either the $z$- (out of plane canting) or the $\perp\mathbf{N}$-direction (in-plane canting), see sketches in Fig.\ref{fig:mn2au_epC_para_zCant_Asymm}. As in the case of the anomalous Hall effect\cite{kipp2021chiral}, we expect the symmetric, crystal photocurrents to closely follow the properties of collinear photocurrents studied above, at least for small canting angles, while marking the sensitive to the sense of chirality chiral photocurrents as a type of currents which are characteristic of canted systems. 
Practically, we realize  the effect of canting by applying an external exchange term to the Wannier interpolated Hamiltonian, as described in Ref.\onlinecite{kipp2021chiral}. Unless stated otherwise, we apply an exchange field $B$ of 200\,meV, keeping in mind that the field with magnitude of $\pm$100\,meV results in about $\pm2^\circ$ canting of  Mn moments. 
We achieve the in-plane (IP) and out of plane (OOP) canting by applying the positive ($+\theta$) and negative ($-\theta$) exchange field perpendicular to the N\'eel vector in-plane, and along  $z$, respectively.

%
We discuss in detail the case of the out of plane canting, presenting in Fig.\ref{fig:mn2au_epC_para_zCant_Asymm} the results for $J^{(s/a)}_{\perp\mathbf{N}}$ (left) and $J^{(s/a)}_{\Vert\mathbf{N}}$ (right) components as defined in Eq.\eqref{eq:symmAsymm_epC} for a 200\,meV exchange field applied along  $z$. 
We observe that the symmetric perpendicular current, $J^{(s)}_{\perp\mathbf{N}}$,  Fig.\ref{fig:mn2au_epC_para_zCant_Asymm}(a), arises in response to linearly polarized light and shows an almost identical frequency dependence as the corresponding photocurrent for the collinear spin configuration shown in Fig.\ref{fig:mn2au_110_condTens_vs_hw}.  The symmetric parallel current, $J^{(s)}_{\parallel\mathbf{N}}$,  Fig.\ref{fig:mn2au_epC_para_zCant_Asymm}(b), responds only to light circularly polarized in the $xy$-plane, which is also the case for the collinear scenario. We have checked that for all considered responses the expectation that the symmetric responses would closely resemble the uncanted ones, generally holds true very well, and we thus discuss in the following only the chiral contributions.

The chiral photocurrent $J^{(a)}_{\perp\mathbf{N}}$ shown in Fig.\ref{fig:mn2au_epC_para_zCant_Asymm}(a) is driven by $xy$-circularly polarized light and is, in contrast to the symmetric component, helicity-switchable, with the largest amplitude $J^{(a)}_{\perp\mathbf{N}}\approx\pm37\times10^{10}$\,A/m$^2$ at the laser frequency of $\hbar\omega=2.0$\,eV. The chiral photocurrent flowing parallel to the N\'eel vector, Fig.\ref{fig:mn2au_epC_para_zCant_Asymm}(b), is comparable in amplitude with $J^{(a)}_{\parallel\mathbf{N}}\approx28\times10^{10}$\,A/m$^2$  at $\hbar\omega=2.8$\,eV. In contrast to the symmetric current for this component,  $J^{(a)}_{\parallel\mathbf{N}}$ is driven by linearly polarized light and its circular part is given by an average over linear components. 
At finite frequencies both chiral photocurrents are comparable in magnitude to their respective crystal counterparts. This is unexpected given a small degree of canting of the moments, however, this observation falls into the philosophy of generally very prominent chiral response in AFMs, also observed for the anomalous Hall effect. \cite{kipp2021chiral} 

We note that due to symmetry chiral charge photocurrents respond to the same type of the field  as the collinear spin photocurrents, if the spin polarization of the spin photocurrents is along the direction of the applied canting field (for example $J^{(a)}_{\parallel\mathbf{N}}$ and $Q^{z}_{\parallel\mathbf{N}}$ both show LPGE-like responses). Interestingly, chiral photocurrent response to linearly polarized light is only present if the exchange field is applied out of the plane. In case of the in-plane canting  sizable chiral photocurrents are also present, however, only in response to circularly polarized light. The presence of chiral photocurrents in response to linearly polarized light might therefore be utilized as a proxy for an out-of plane switching path accompanied by canting. 
\subsection{Chiral IFE and chiral spin photocurrents}\label{section6b}
%
%
%
%
%
%
%
%


Motivated in part by our observation of strong chiral photocurrents of charge, and given a strong recent interest in the physics and properties of spin currents in chiral systems\cite{kimata2019magnetic,nan2020controlling,masuda2022chirality,kimata2019magnetic,go2022noncollinear}, we finally discuss the modifications to IFE and spin photocurrents brought by canting in Mn$_2$Au. Analogously to 
Eq.\eqref{eq:symmAsymm_epC} we define the symmetric (crystal) and antisymmetric (chiral) components of the laser induced spin density and spin photocurrent with respect to a canting by angle $\theta$ as
\begin{equation}
    \begin{aligned}
        \delta S^{a/s} &=   \frac{\delta S^+( + \theta ) \pm \delta S^+(- \theta)}{2} \\
        \mathrm{Q}^{a/s} &=   \frac{\mathrm{Q}( + \theta ) \pm \mathrm{Q}(- \theta)}{2}. \\
    \end{aligned}
     \label{eq:symmAsymm_spD}
\end{equation}
Note that we consider only the chiral component of the total photo-induced spin density $\delta S^+$ and spin photocurrent $Q$, as we did not find either qualitative changes or significant magnitude in other components upon canting.

\begin{figure} 
    \centering
    \includegraphics[width=0.95\columnwidth]{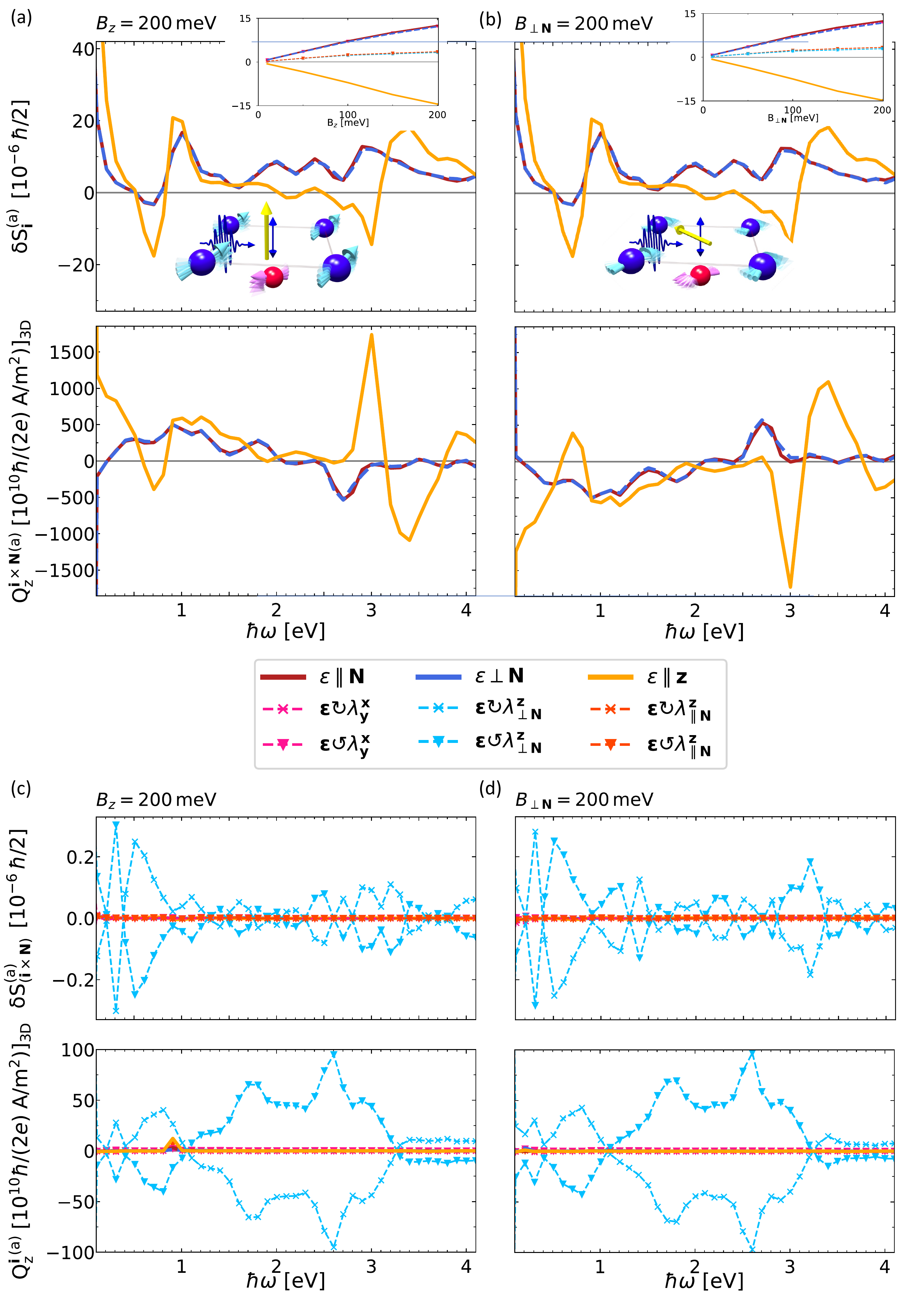}
    \caption[Laser induced SGE in canted Mn$_2$Au]{Antisymmetric spin responses as a function of light frequency for (a,c) in-plane canting (IP, exchange field $B$ along $\mathbf{i}\perp \mathbf{N}$) and (b,d) out of plane canting (OOP, exchange field $B$ along $\mathbf{i}\Vert z$). Shown are the laser-induced nonequilibrium spin density and the asssociated spin photocurrent. (a,b) The exchange field applied along $\mathbf{i}$ drives a LPGE-like chiral spin density parallel to the field. The associated LPGE chiral spin photocurrent flows along $z$ and it is polarized along $\mathbf{i}\times{\mathbf{N}}$. Note the colossal magnitude of the chiral spin currents. The insets show the dependence of the antisymmetric spin density on the strength of the applied exchange field for various laser polarizations (at $\hbar\omega=3.0$\,eV). 
    (c,d) The exchange field also drives a CPGE-like chiral spin density along $\mathbf{i}\times\mathbf{N}$ direction. The associated CPGE-like spin photocurrent propagates along $z$ and it is polarized perpendicular to the induced spin density along $(\mathbf{i}\times\mathbf{N})\times\mathbf{N}=\mathbf{i}$.  The broadening for all calculations has been taken at the value of 25\,meV.}
    \label{fig:mn2au_canted_SGE_vs_hw}
\end{figure}

In Fig.\ref{fig:mn2au_canted_SGE_vs_hw} we present the frequency dependence of the  LPGE-like (a-b) and CPGE-like (c-d) chiral nonequilibrium spin density and spin current for the IP (a,c) and OOP (b,d) canting.  
%
%
%
Displayed in Fig.\ref{fig:mn2au_canted_SGE_vs_hw}(a,b) is the chiral spin density induced along the direction $\mathbf{i}$ of the applied IP and OOP exchange field. The induced spin density is clearly LPGE-like with sizeable magnitude  for all orientations of the linearly polarized light. Comparing the spin density response in (a) and (b) we realize that induced $\delta S^a_{\mathbf{i}}$ does not at all depend on the direction of the applied exchange field. Moreover, the response is perfectly isotropic with respect to the rotation of the polarization of light in the plane.
At the frequency of $\hbar\omega=1.6\,$eV the spin density is suppressed for all three linear independent  polarizations. Up to this frequency the two in plane and $z$-polarized signals have the same sign and frequency dependence, with the IP-polarized response smaller in magnitude. At frequencies 
above $\hbar\omega=2.0\,$eV the OOP response becomes negative while the IP responses remain positive. For example at $\hbar\omega=2.9\,$eV the spin density can be switched between the values of $\delta S^{(a)}_{\perp\mathbf{N}}\approx\pm 10\times10^{-6} \tfrac{\hbar}{2}$ when rotating the linear polarization from within the $xy$-plane into the $z$-direction.
The discussed chiral spin density can be seen as a result of an action of an exchange field on the staggered spin density generated along $\mathbf{N}$, $\delta S^-_{\Vert \mathbf{N}}$, which gets canted,
resulting in an effective ferromagnetic moment and associated chiral spin density pointing along the exchange field. This spin density response 
can effectively drive a spin-polarized current via the spin galvanic effect \cite{ganichev2002spin}. Indeed, we observe a generation of colossal LPGE-like spin photocurrents 
which propagate along $z$ with spin-polarization along $\mathbf{i}\times\mathbf{N}$, see lower part of Fig.\ref{fig:mn2au_canted_SGE_vs_hw}(a,b), with the magnitude which is larger by an order of magnitude as compared to collinear spin photocurrents, discussed above, reaching as much as $ 1760\times10^{10}\,[\hbar/(2e)\,\mathrm{A/m^2}]$ for light polarized along $z$ at $\hbar\omega=3.0$\,eV. Interestingly, while the LPGE chiral photocurrents of charge are significantly suppressed for the IP canting (not shown), the chiral spin photocurrents are
found to correlate directly
with the chiral spin density, both in terms of the frequency dependence and light polarization dependence, as well as in their ``insensitivity'' to the direction of the applied canting field.  
As the insets show, the chiral laser-induced spin density scales linearly with the applied exchange field up to the field strength of 200\,meV, and we also confirm that the linear behavior in this range is preserved by other computed quantities. This implies that even at very small canting angles below one degree, the photo-induced chiral spin currents will be very large in overall magnitude, and can serve as the markers of both the effect of canting as well as its strength.

 


Besides the LPGE-like responses shown in Fig.\ref{fig:mn2au_canted_SGE_vs_hw}(a,b), a helicity-switchable chiral CPGE-like photo-induced nonequilibrium spin density is induced  along $\mathbf{i}\times\mathbf{N}$, accompanied by an associated spin-polarized photocurrent flowing along $z$ with spins along $\mathbf{i}$, as shown in Fig.\ref{fig:mn2au_canted_SGE_vs_hw}(c,d). However the predicted magnitude of the effects is two orders of magnitude smaller when compared to the LPGE-like responses, being overall comparable to the range of values for the collinear case. Here, we also do not observe any correlation in frequency dependence between the photo-induced spin density and the spin currents.



\section{Conclusions}

In our work, using ab-initio Keldysh formalism, we have scrutinized the response properties of Mn$_2$Au subject to linearly and circularly polarized laser light. We explicitly computed the laser-induced photocurrents of charge and spin, and laser-induced spin-polarization. We found that the magnitude of the computed effects is sizeable, and their nature is of purely magnetic origin. The extreme sensitivity of the effects to the sense of light polarization, direction of the N\'eel vector, light frequency and quasiparticle lifetime was observed. The diversity of observed response suggests that the dynamical properties of staggered magnetization in this material can be detected by tracking the dynamics of the photocurrents. We have uncovered and studied the emergence of the photospin Hall effect in Mn$_2$Au, which governs the generation of transverse photocurrents of spin in response to generated charge photocurrents in this system. Finally, we predict that colossal chiral spin photocurrents can be generated even upon a very small canting of staggered moments, which suggests that antiferromagnetic dynamics  accompanied by intrinsic reorganization of the magnetic order may find a prominent place in optospintronics applications as a strong source of optically generated spin currents.

\begin{acknowledgments}
This work was supported by the Deutsche Forschungsgemeinschaft (DFG, German Research Foundation) $-$ TRR 173/2 $-$ 268565370 (project A11), TRR 288 – 422213477 (project B06), and the Sino-German research project DISTOMAT (MO 1731/10-1). This project has received funding from the European Union’s Horizon 2020 research and innovation programme under the Marie Skłodowska-Curie grant agreement No 861300.
\end{acknowledgments}

\bibliography{references}

\end{document}